\documentclass[journal]{IEEEtran}
\usepackage{amsmath} 
\usepackage{amssymb}
\usepackage{amsfonts}
\usepackage{amsthm}
\allowdisplaybreaks[4]
\usepackage{algorithmic}
\usepackage{algorithm}
\usepackage{gensymb}
\usepackage{array}
\usepackage{booktabs}
\usepackage[caption=false,font=normalsize,labelfont=sf,textfont=sf]{subfig}
\usepackage{textcomp}
\usepackage{color}
\usepackage{bm}
\usepackage{stfloats}
\usepackage{caption}
\usepackage{url}
\usepackage{verbatim}
\usepackage{graphicx}
\usepackage{diagbox}
\usepackage{epstopdf}
\usepackage{cite}

\newtheorem{theorem}{Theorem}
\begin{document}

\title{Beamforming Design and Multi-User Scheduling in Transmissive RIS Enabled Distributed Cooperative ISAC Networks with RSMA}

\author{Ziwei~Liu,~Wen~Chen,~Qingqing~Wu,~Zhendong~Li,~Qiong~Wu,~Nan Cheng,~and~Jun Li
	

	\thanks{Z. Liu, W. Chen, and Q. Wu are with the Department of Electronic Engineering, Shanghai Jiao Tong University, Shanghai 200240, China (e-mail: ziweiliu@sjtu.edu.cn; wenchen@sjtu.edu.cn; qingqingwu@sjtu.edu.cn).}
	\thanks{Z. Li is with the School of Information and Communication Engineering, Xi'an Jiaotong University, Xi'an 710049, China (e-mail: lizhendong@xjtu.edu.cn). }	
		\thanks{Q. Wu is with the School of Internet of Things Engineering, Jiangnan
		University, Wuxi 214122, China (e-mail: qiongwu@jiangnan.edu.cn). }
	\thanks{N. Cheng is with the State Key Lab. of ISN and School of Telecommunications Engineering, Xidian University, Xi'an 710071, China (e-mail: dr.nan.cheng@ieee.org).	}
	\thanks{J. Li is with the School of Information Science and Engineering, Southeast University, Nanjing, 210096, China. (email: jleesr80@gmail.com).}

}

%

\maketitle

\begin{abstract}
In this paper, we propose a novel transmissive reconfigurable intelligent surface (TRIS) transceiver-empowered
distributed cooperative integrated sensing and communication (ISAC) network to enhance coverage as well as to enhance wireless environment understanding. Based on the network requirements, the users are categorized into cooperative users (CUEs) and destination users (DUEs), and the CUEs utilize their own resources to serve the DUEs. To realize cooperation, we implement rate-splitting multiple access (RSMA) at the base
station (BS), where the common stream is decoded and reencoded at the CUEs and forwarded to the DUEs, while the
private stream satisfies the CUEs’ own communication requirements. We construct an optimization problem with maximum minimum radar mutual information (RMI) as the objective function to optimize the BS beamforming matrix, the CUE beamforming matrices, the common stream rate vectors, and the user scheduling vectors. Due to the coupling of the optimization variables and non-convex operation, the proposed problem is a
non-convex optimization problem that cannot be solved directly. To address the above challenges, we adopt a consensus alternating direction method of multipliers (ADMM) framework to decouple the optimization variables and solve it. Specifically, the problem is decoupled into multiple subproblems and solved by iterative
optimization independently until overall convergence is achieved. Finally, numerical results validate the superiority of the proposed scheme in terms of improving communication sum-rate and RMI, and greatly reduce the algorithm complexity.
\end{abstract}

\section{Introduction}
\IEEEPARstart{O}{n} contemporary era, the tremendous advancement of science and technology has promoted the deep fusion of information technology and communication networks. A notable feature in this process is the emergence of multiagent to achieve information sharing, environment sensing and cooperative decision making in networks. This necessitates exploring how cooperative sensing can be utilized to improve system intelligence, enhance environment understanding, and enable efficient communication\cite{10273396,9810792,9585321}. 

In traditional sensing networks, agents usually rely on their own sensors for information acquisition, however, in complex and dynamic environments, it is incapable of ensuring the accuracy and comprehensiveness of such an isolated sensing approach. Through cooperative integration, multi-agents can share their own sensory information to jointly construct a more accurate and complete environment model \cite{8910629,10478321}. Consequently, better network decision-making, target detection, and wireless environment understanding can be realized \cite{9170532,10666007,10561589}. In \cite{9170532}, cooperative spectrum sensing in cognitive radio networks is discussed. The sub-transmitters send their local sensing decisions to a fusion center, which makes the better final decision on the spectrum state by means of fusion rules. In \cite{10666007}, the problem of sensor-assisted cooperative spectrum sensing for unmanned aerial vehicles is investigated. The sensor identifies available spectrum resources and opportunistically accesses channels that are underutilized by the primary user, thereby achieving cooperative gain and a significant advantage in detection performance. In \cite{10561589}, a cooperative sensing scheme with multiple reference signals is designed, and it is verified that multiple reference signals can improve the sensing accuracy more effectively than single reference signal. The process of cooperative sensing relies on the sharing of information, for which communication technologies provide a solution to facilitate network performance.

In cooperative information sharing, communication technology plays a key bridging role. Effective cooperative communication mechanisms ensure timely sharing and seamless integration of sensory information and have great potential to extend communications coverage and improve spectrum efficiency \cite{9994493,10478825,10380500}. In \cite{9994493}, the design of the multiuser cooperative communication system is studied. Through verification, the proposed multiuser cooperative system improves the transmission performance of the system. In \cite{10478825}, the subject of cooperative communication employing rate-splitting with a certain probability of channel obstruction is investigated, and the results show that the cooperative approach enhances the coverage. In \cite{10380500}, multiple satellites cooperating to provide services to multiple ground users are considered. An efficient multi-satellite cooperative transmission framework is proposed, and the results show the superiority of the proposed scheme in terms of spectral efficiency. Moreover, the target echo results from the communication transmitters can also be used to improve the sensing performance. However, since communication and sensing involve different performance metrics, design standards, and waveforms, this introduces issues such as interference and impacts on the integrated design.

The emergence of integrated sensing and communication (ISAC) technology, which enables communication signals andsensing signals to be realized through integrated waveforms, provides a new solution for future network design \cite{10077114}. A non-orthogonal multiple access-assisted ISAC system based on sensing scheduling is investigated in \cite{10129092}, aiming to maximize the sensing efficiency of the ISAC system while meeting the communication quality requirements, and achieve better communication and sensing performance. In \cite{10032141}, a cooperative ISAC system is investigated that employs a rate-splitting multiple access (RSMA) transmission scheme for advanced interference management and a direct localization sensing scheme to provide highly accurate positioning services. Numerical results show that the optimal performance is achieved by providing greater degrees of freedom for the design of the rate-splitting strategy. Moreover, the performance of the cooperative radar and communication system is improved over the non-cooperative and independently operating system model \cite{9580023}. However, the cooperative approach still leaves a number of problems to be solved.

In particular, coverage and energy consumption need to be considered, as future ISAC networks have high signal
frequencies, signals are easily blocked, and the introduction of cooperative BS will inevitably bring about an increase in energy consumption. Reconfigurable intelligent surface (RIS) is a promising option as a low-power device for enhanced coverage \cite{9610992}. In \cite{10058895}, a RIS-assisted cell-free massive multiple-input multiple-output network is investigated and the cooperative beamforming is designed for coverage
enhancement. The RIS-assisted ISAC network considering blocking effect is investigated in \cite{10695883}. By utilizing stochastic geometry to study the favorable impact of RIS on coverage, the scheme achieves a significant increase in the joint coverage of ISAC performance. Furthermore, a transmissive RIS (TRIS) transceiver is investigated in \cite{10680462,Liu2023}, capable of delivering higher system performance gains in the form of low power consumption, which is more suitable for reducing power consumption in cooperative networks. Moreover, TRIS transceivers outperform traditional transceivers in improving the network’s sensing and communication performance, and consume significantly less energy \cite{10522473,liu2024enhancingrobustnesssecurityisac}.

Overall, users in future networks will be involved in the collection, sharing and forwarding of network information, which faces many problems such as user scheduling, resource allocation and energy consumption, and poses new challenges for future network design. However many existing works rarely synthesize these considerations and little research has been conducted on cooperative ISAC. In addition, interference management is an open issue, and flexible interference management scheme such as RSMA is a new solution for future network development. In this paper, a cooperative ISAC network architecture is proposed, in which cooperative users are equipped with low-power consumption and low-cost TRIS transceivers, and RSMA is used for message segmentation and delivery. Based on the scenario, a distributed cooperative ISAC optimization problem is constructed with many constraints such as energy consumption, user scheduling, etc. The main contributions of this paper are as follows:
\begin{itemize}
\item[$\bullet$] We establish a cooperative ISAC network architecture. In this architecture, it is divided into feasible and blocked areas based on the coverage of the base station (BS), where the cooperative users (CUEs) are distributed in the feasible areas and the destination users (DUEs) are distributed in the blocked areas. The common stream of RSMA is decoded and recoded at the CUEs for messaging and environment sensing of the DUEs, and the private stream is decoded by the CUEs to satisfy their own communication demands. In this process, CUEs are equipped with TRIS transceivers with low power consumption due to energy consumption considerations. To the best of our knowledge, there are few researches on utilizing the common stream and the private stream of RSMA, respectively, to satisfy the needs of different users in a cooperative network.
\item[$\bullet$] Based on the architectural design, we propose a distributed cooperative ISAC algorithm, which is based on the consensus alternating direction method of multipliers (ADMM) framework. The original problem is divided into independent subproblems for solving, and the optimal closed-form solutions can be obtained iteratively by constructing the Lagrangian functions, which in turn can be computed in a distributed manner at different CUEs, greatly reducing the complexity of the algorithm. Moreover, since the user scheduling is an integer optimization problem, we first relax it into a continuous variable, and then propose a threshold-based recovery algorithm for the user scheduling variable.
\item[$\bullet$] We evaluate the performance of the proposed distributed cooperative ISAC scheme with a new TRIS transceiver, and numerical results show that the proposed scheme improves the sensing performance by at least 40.48\%, the communication performance by at least 19.47\%, and reduces the computational complexity by up to 92.7\% over the baseline schemes. Further, the proposed algorithm still has relatively satisfactory sensing and communication performance in the case of traditional transceivers, which confirms the superiority of the proposed algorithm.
\end{itemize}

\emph{Notations}: Scalars are denoted by lower-case letters, while vectors and matrices are represented by bold lower-case letters and bold upper-case letters, respectively. $|x|$ denotes the absolute value of a complex-valued scalar $x$, ${x^ * }$ denotes the conjugate operation, and $\left\| \bf x \right\|$ denotes the Euclidean norm of a complex-valued vector $\bf x$. For a square matrix ${\bf{X}}$, ${\rm{tr}}\left( {\bf{X}} \right)$, ${\rm{rank}}\left( {\bf{X}} \right)$, ${{\bf{X}}^H}$, ${\left[ {\bf{X}} \right]_{m,n}}$ and $\left\| {\bf{X}} \right\|$ denote its trace, rank, conjugate transpose, ${m,n}$-th entry, and matrix norm, respectively. ${\bf{X}} \succeq 0$ represents that ${\bf{X}}$ is a positive semidefinite matrix. In addition, ${\mathbb{C}^{M \times N}}$ denotes the space of ${M \times N}$ complex matrices. $j$ denotes the imaginary element, i.e., $j^2 = -1$. The distribution of a circularly symmetric complex Gaussian (CSCG) random vector with mean $\mu $ and variance $\sigma^2$ is denoted by ${\cal C}{\cal N}\left( {{\mu},\sigma^2} \right)$ and $ \sim $ stands for ‘distributed as’. ${\bf{A}} \otimes {\bf{B}}$ represents the Kronecker product of matrices ${\bf{A}}$ and ${\bf{B}}$. ${\bf{A}} \circ {\bf{B}}$ denotes the Hadamard product of matrices ${\bf{A}}$ and ${\bf{B}}$.

\section{System Model}
In this section, we consider a multiuser multiple input single output (MISO) cooperative communication and sensing (C\&S) scenario in the millimeter-wave band. The network consists of a TRIS empowered BS, $K$ CUEs, and $M$ single antenna DUEs, as shown in Fig. 1. Due to the millimeter-wave band, scatterers and obstacles in the wireless environment will negatively affect the coverage of the BS, which can be divided into feasible and blocked regions with different radiicentered on the BS according to the actual environment. Among them, the CUEs are the users in the feasible area, which utilize their own resources as cooperative decode-and-forward (DF) relay toprovide C\&S services for all DUEs in the blocked area under the condition of meeting their own communication requirements. Moreover, information sharing among CUEs through BS and the space and time registration scheme can be referred to\cite{9800700}.

\begin{figure}[H]
		\centerline{\includegraphics[width=8.0cm]{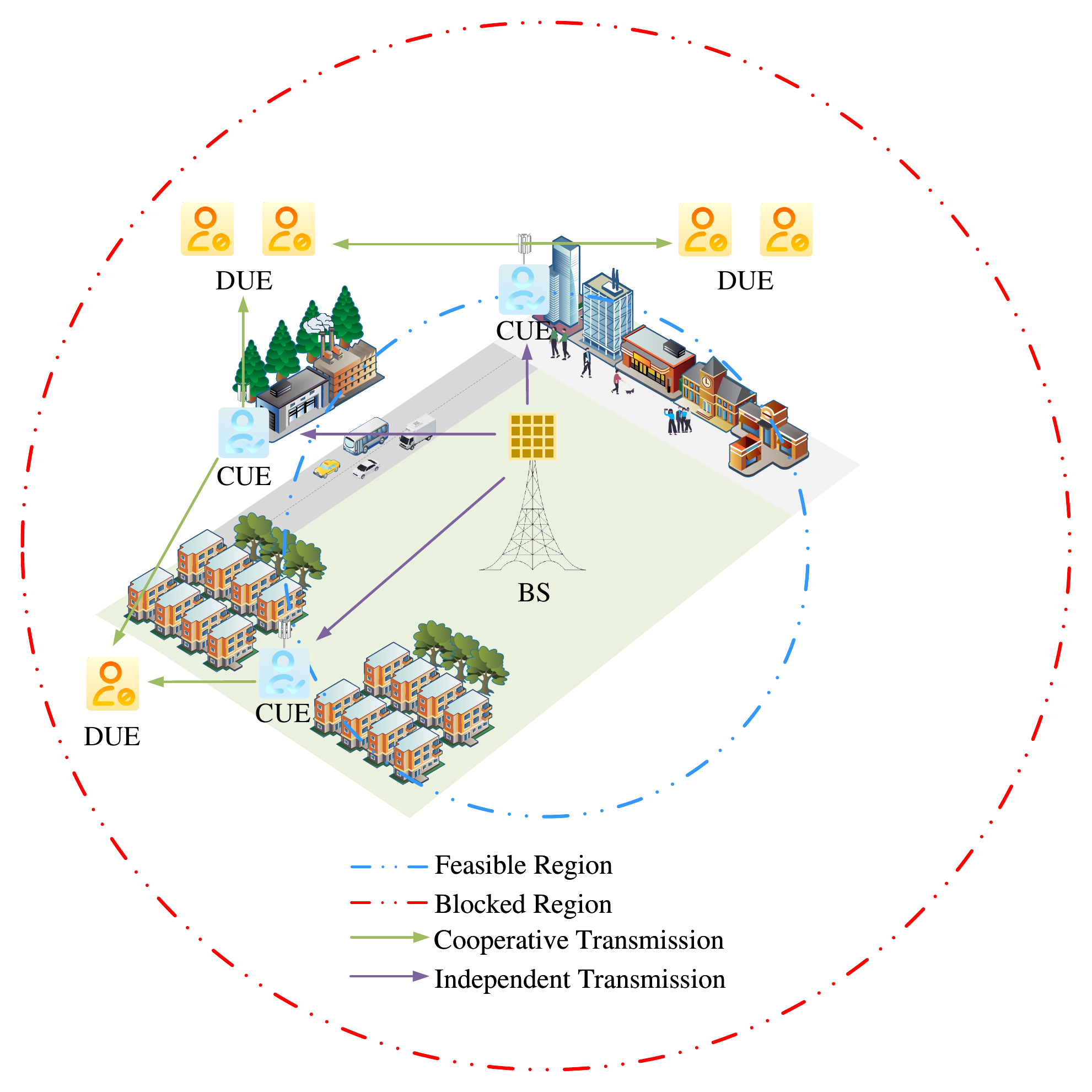}}
	\caption{Cooperative ISAC networks.}
\end{figure}
\subsection{Characteristics of TRIS}
In future wireless networks, each user may become a potential collaborative access point that utilizes its own resources to enhance the throughput of the entire network as well as its ability to access information. It is expected that these access points will utilize their own resources as minimally as possible, while meeting their own service requirements, therefore TRIS transceiver is aviable option.

TRIS transceiver is a low power consumption multistream communication scheme, ideal for dynamic wireless environments and low power consumption requirements, which consists of a TRIS with $N=N_r\times N_c$ reconfigurable elements arranged in a uniform planar array (UPA) pattern, a horn antenna and a controller\footnote{This structure has been investigated in many literatures \cite{10740042,10242373,10711860}, where TRIS acts as a loader of information, which utilizes the control signals generated by the time modulation array (TMA) to load precoding bit stream information onto different TRIS elements, the carrier penetrates the elements and carries the information to achieve direct digital modulation, thus realizing the functionality of multiple antennas with lower energy consumption}. Due to the characteristics of the control signals generated by the TMA and the maximum energy loading demands of the TRIS elements, the following constraint is required for the precoding matrix\cite{10711860}
\begin{equation}
	\begin{bmatrix}\mathbf{W}\mathbf{W}^H\end{bmatrix}_{nn}\leq P_t,\forall n,
\end{equation}
where $\bf W$ denotes precoding matrix and $P_t$ denotes the maximum available power of each TRIS element. In order to satisfy the demands and convenience of the subsequent algorithm design, constraint (1) is equivalently transformed into the following form
\begin{equation}
	\left\|\mathrm{vec}\left(\mathbf{W}\right)\circ\mathbf{c}_n\right\|^2\leq P_t,\forall n,
\end{equation}
where ${\rm vec}\left(  \cdot  \right)$ denotes the vectorization of $\bf W$ by columns. ${\bf c}_n$ denotes the position index, which takes the value 1 at positions $(n,1),\cdots,(n,K)$ and 0 for the rest, and can be expressed as
\begin{equation}
{{\bf{c}}_n} = \left( {0, \cdots ,\overbrace {\mathop 1\limits_{\left( {n,1} \right)} ,0, \cdots \mathop 1\limits_{\left( {n,2} \right)} }^N,0, \cdots ,\mathop 1\limits_{\left( {n,K} \right)} } \right),\forall n.
\end{equation}

\subsection{Channel Model} 
In this paper, CUEs and DUEs have different service requirements and we first consider the communication channel for CUE. Since the CUE is in the feasible region of the BS, its channel is highly directional and the scattering of the signal is small, so we consider the line-of-sight (LoS) channel ${{\bf{h}}_{k}} \in {\mathbb{C}^{N \times 1}}$ whose small-scale fading effect is negligible and can be expressed as
\begin{equation}
	{{\bf{h}}_{k}} = {\xi _{k}}{\left[ {{e^{ - j\pi {{\sin {\theta _{k}}\cos {\varphi _{k}}}}{{\bf{n}}_r}}}} \right]} \otimes {\left[ {{e^{ - j2\pi{{\sin {\theta _{k}}\sin {\varphi _{k}}}}{{\bf{n}}_c}}}} \right]},\forall k,\label{h}
\end{equation}
where ${\xi _{k}={{\lambda _c}}/{{4\pi {d_{k}}}}}$ denotes the path loss, ${\lambda _c}$ represents the carrier wavelength, and ${d_{k}}$ represents the distance between the BS and the CUE.  ${{\bf{n}}_r} = \left[ {0,1, \cdots ,{N_r} - 1} \right]^T$, ${{\bf{n}}_c} = \left[ {0,1, \cdots ,{N_c} - 1} \right]^T$, and $\left( {{n_r},{n_c}} \right)$ denotes the element position index of the RIS.  ${\varphi _{k}}$ and ${\theta _{k}}$ denote the azimuth and pitch angles between the BS and the CUE, respectively.

Since scattering and occlusion exist in the obstructed area, the channel ${{\bf{g}}_{k,m}} \in {\mathbb{C}^{N \times 1}}$ between the $k$-th CUE and the $m$-th DUE is modeled as a Rician fading channel as follows
\begin{equation}
	{{\bf{g}}_{k,m}} = {\xi _{k,m}}\left( {\sqrt {\frac{{{\kappa _v}}}{{{\kappa _v} + 1}}} {\overline {\bf{g}} _{k,m}} + \sqrt {\frac{1}{{{\kappa _v} + 1}}} {\underline {\bf{g}}_{k,m}}} \right),\forall k,m,\label{g}
\end{equation}
where ${\kappa _v}$ represents the Rician factor. The expressions for LoS channel ${\overline {\bf{g}} _{k,m}}$ and the path loss ${\xi _{k,m}}$ are similar to those defined in Eq. (\ref{h}), and the non-line-of-sight (NLoS) channel ${\underline {\bf{g}}_{k,m}}$ obeys a CSCG distribution, i.e., ${\underline {\bf{g}}_{k,m}}\ \sim {\cal C}{\cal N}\left( {0,{{\bf{I}}_N}} \right)$. 
Subsequently, in order to better serve the DUEs and have a better understanding of the wireless environment, the round-trip sensing channel ${{\bf z}_{k,m,u}}\in {\mathbb{C}^{N \times 1}}$ is modeled as
\begin{equation}
{{\bf z}_{k,m,u}}	= \alpha_{k,m,u}e^{-j2\pi d_{u,m}/\lambda_c}{\bf a}\left(\theta_{k,m},\varphi_{k,m}\right),\forall k,m,u,
\end{equation}
where ${\bf a}\left(\theta_{k,m},\varphi_{k,m}\right)$ denotes the transmit steering vector between the $k$-th CUE and the $m$-th DUE, $e^{-j2\pi d_{u,m}/\lambda_c}$ denotes the channel from the $m$-th DUE to the receiving antenna of the $u$-th CUE, and $\alpha_{k,m,u}$ denotes the complex reflection coefficient. ${\alpha _{k,m,u}} \sim {\cal C}{\cal N}\left( {0,\sigma _{{\alpha _{k,m,u}}}^2} \right)$, $\sigma _{{\alpha _{k,m,u}}}^2 = {S_{RCS}} {\lambda_c}^2 / {(4\pi)^3d_{k,m}^2d_{u,m}^2}$, and ${S_{RCS}}$ denotes the RCS of target\cite{ding}. The transmit steering vector can be expressed as 
\begin{equation}
		\begin{split}
	{\bf a}\left(\theta_{k,m},\varphi_{k,m}\right)&={\left[ {{e^{ - j\pi {{\sin {\theta _{k,m}}\cos {\varphi _{k,m}}}}{{\bf{n}}_r}}}} \right]}\\ &~~~~~~~\otimes {\left[ {{e^{ - j2\pi{{\sin {\theta _{k,m}}\sin {\varphi _{k,m}}}}{{\bf{n}}_c}}}} \right]},\forall k,m.
		\end{split}
\end{equation}
The channels ${\bf h}_k$, ${\bf g}_{k,m}$ and ${\bf z}_{k,m,u}$ described above can be obtained by channel estimation\cite{8835503,9839429,9732214}.

\begin{figure*}
	{\centerline{\includegraphics[width=18cm]{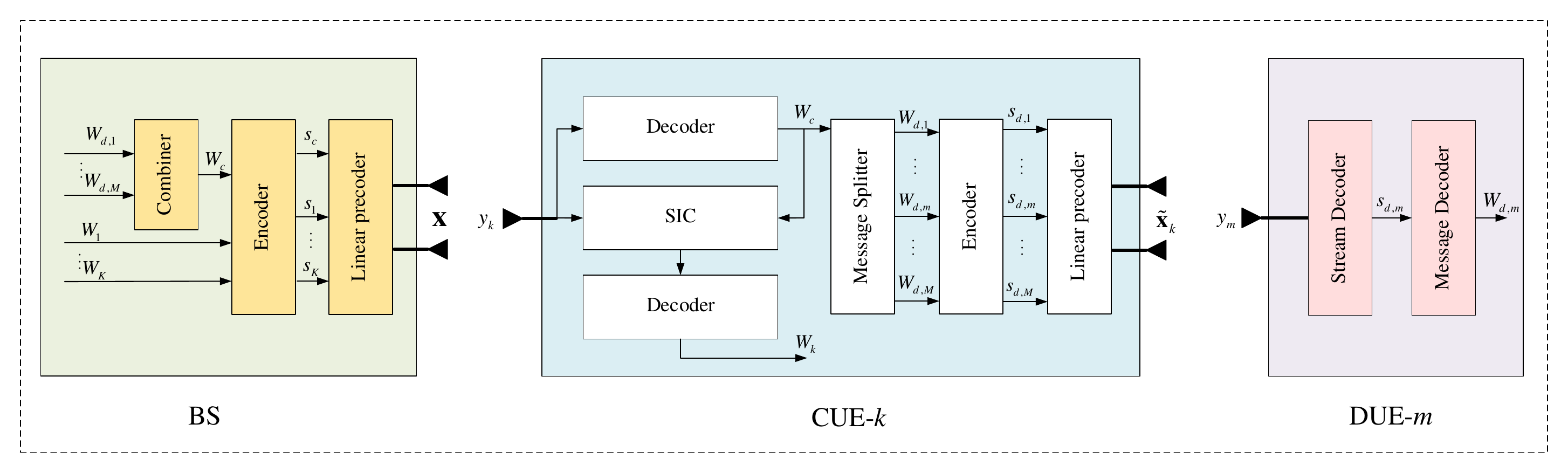}}}
	\caption{Cooperative rate-splitting framework.}
\end{figure*}
\subsection{Signal Model}

For cooperative communication, rate splitting signal is implemented. The signal sent by the BS can be expressed as
\begin{equation}
	\begin{split}
		{\bf{x}} = {{\bf{W}}{\bf{s}}} = {{{\bf{w}}_c}{s_c}} + {\sum\nolimits_{k = 1}^K {{{\bf{w}}_k}{s_k}} },
	\end{split}
\end{equation}
where ${\bf{W}} = \left[ {{{\bf{w}}_c},{{\bf{w}}_1}, {{\bf{w}}_2},\cdots ,{{\bf{w}}_K	}} \right] \in \mathbb{C}{^{N \times \left( {K + 1} \right)}}$ denotes the linear beamforming matrix at BS, and ${\bf{s}} = {\left[ {{s_c},{s_1}, {s_2}, \cdots ,{s_K}} \right]^T} \in \mathbb{C}{^{\left( {K + 1} \right) \times 1}}$ represents the encoded signal, which is a wide stationary process in the time domain, statistically independent and with zero mean. 

In this paper, the information for all DUEs $\left\{ {W_{d,1}, \cdots ,W_{d,M}} \right\}$ are combined as $W_c$ and jointly encoded as a common stream $s_c$ and the information for all CUEs $\left\{ {{W_{1}}, \cdots ,{W_{K}}} \right\}$ are independently encoded as private streams $\left\{{s_1},\cdots ,{s_K}\right\}$. The BS sends the encoded signal to CUEs and the common stream is decoded at the CUEs, then the common stream is removed from the received signal and thus the private stream can be decoded in the following. Then the common stream is split into $\left\{ {W_{d,1}, \cdots ,W_{d,M}} \right\}$ and encoded into ${\bf s}_d =\left\{{s_{d,1}},\cdots ,{s_{d,M}}\right\}$. Finally, the recoded signal $\tilde{\bf x}_k$ is sent to the DUE after linear encoding. The encoding and decoding process is shown in Fig. 2. The recoded signal $\tilde{\bf x}_k$ can be expressed as
\begin{equation}
	\begin{split}
		\tilde{\bf x}_k={\bf F}_k{\bf s}_d=\sum\nolimits_{m = 1}^{M}{{{\bf{f}}_{k,m}}}{{s_{d,m}}},\forall k,
	\end{split}
\end{equation}
where ${\bf F}_k=\left[{{{\bf{f}}_{k,1}}},{{{\bf{f}}_{k,2}}},\cdots,{{{\bf{f}}_{k,M}}}\right]\in \mathbb{C}^{N\times M}$ denotes the $k$-th CUE beamforming matrix.

The signal received by the $k$-th CUE can be expressed as
\begin{equation}
	{y_k} = \sum\nolimits_{i = 1}^{\cal \hat K}{\widetilde {\bf{h}}_k}^H\left({\rm vec}\left({\bf W}\right) \circ {\bf a}_i\right){{s_i}}+ {n_k},\forall k,
\end{equation}
where ${n_k} \sim {\cal C}{\cal N}\left( {0,{\sigma _{{n_k}}^2}} \right)$ denotes the Gaussian white noise at the CUE and ${\cal \hat K}=\left\{c,1,...,K\right\}$. ${\widetilde {\bf{h}}_k}=\left[{{\bf{h}}_k};{{\bf{h}}_k};\cdots,{{\bf{h}}_k}\right]\in \mathbb{C}^{N(K+1)\times 1}$ denotes the $(K+1)$-fold extension to channel ${{\bf{h}}_k}$. Similarly, ${\rm vec}\left({\bf W}\right)=\left(w_{c,1},\cdots,w_{c,N},w_{1,1},\cdots,w_{1,N},\cdots,w_{K,1},\cdots,w_{K,N}\right)^T\in \mathbb{C}^{N(K+1)\times 1}$ denotes the vectorization of the matrix ${\bf W}$ by columns. ${\bf a}_i\in\mathbb{C}^{N(K+1)\times1}$ denotes the position index, which has a value of $1$ at positions $(i,1)\sim(i,N)$ and $0$ at the remaining positions, as follows
\begin{equation}
	{\bf a}_i = \left( {0, \cdots ,0,\underbrace {1, \cdots ,1}_{\left(i,1\right) \sim \left(i,N\right)},0, \cdots ,0} \right)^T,\forall i\in {\cal \hat K}.
\end{equation}

As mentioned above, CUEs provide C\&S services to DUEs through cooperation. Therefore, each DUE suffers from interference from neighboring DUEs and distant DUEs. Note that the visible CUEs for each DUE are known, but the specific link between the CUEs and the DUE is uncertain and needs to be determined. Based on this fact, we introduce a discrete variable $\rho_{k,m}$ to indicate whether a link is established or not, where $\rho_{k,m}=1$ if CUE $k$ serves DUE $m$ and $\rho_{k,m}=0$ otherwise. The signal received by the $m$-th DUE can be expressed as
\begin{equation}
	{y_m} = \sum\nolimits_{k = 1}^{K}\sum\nolimits_{j = 1}^{M}\widetilde{\bf{g}}_{k,m}^H\left({\rm vec}\left({\bf F}_{k}\right) \circ {\bf b}_m\circ {\bf p}_k\right){{s_{d,j}}}+ {n_m},\forall m,
\end{equation}
where ${n_m} \sim {\cal C}{\cal N}\left( {0,{\sigma _{{n_m}}^2}} \right)$ denotes the Gaussian white noise at the DUE. ${\widetilde {\bf{g}}_{k,m}}=\left[{{\bf{g}}_{k,m}};{{\bf{g}}_{k.m}};\cdots,{{\bf{g}}_{k,m}}\right]\in \mathbb{C}^{NM\times 1}$ denotes the $M$-fold extension of channel ${{\bf{g}}_{k,m}}$. ${\bf p}_k=\left[{\bm \rho}_{k};{\bm \rho}_{k};\cdots;{\bm \rho}_{k}\right]$ denotes the $N$-fold expansion of the user scheduling variable ${\bm \rho}_{k}=\left(\rho_{k,1},\rho_{k,2},\cdots,\rho_{k,M}\right)^T$. ${\bf b}_m\in \mathbb{C}^{MN\times1}$ denotes the position index, which has a value of $1$ at positions $(m,1)\sim(m,N)$ and $0$ at the remaining positions and is defined similarly to ${\bf a}_i$.

For cooperative sensing, all CUEs transmit a recoded common stream ${\tilde{\bf x}_k}$ for sensing and receive the reflected echoes of all CUE probe signals. The echo signal at the $u$-th CUE\footnote{The $u$-th CUE can receive reflected signals from all CUEs transmitting to all DUEs, but in practice the link is not established due to beam selection and user scheduling as well as the blocking of some reflected echoes, which is accounted for using the variable $\rho_{k,m}$, i.e., indicated by $\rho_{k,m}=0$. } can be expressed as
\begin{equation}
	y_{u} = \sum\nolimits_{k=1}^{K}\sum\nolimits_{m=1}^{M}\widetilde{\bf z}_{k,u}^H\left({\rm vec}\left({\bf F}_{k}\right) \circ {\bf b}_m\circ {\bf p}_k\right)s_{d,j}+n_{u},\forall u,
\end{equation}
where ${n_u} \sim {\cal C}{\cal N}\left( {0,{\sigma _{{n_u}}^2}} \right)$ denotes the Gaussian white noise at the $u$-th CUE and $u\in\left\{1,\cdots,K\right\}$. $\widetilde{\bf z}_{k,u}=\left[{\bf z}_{k,1,u};{\bf z}_{k,2,u};\cdots;{\bf z}_{k,M,u}\right]\in \mathbb{C}^{NM\times 1}$ denotes the stack of $M$ channels ${\bf z}_{k,m,u}$.

\section{Performance Metrics and Problem Formulation}
In this section, we provide the performance metrics for communication and sensing, respectively, and formulate the optimization problem.

\subsection{Performance Metrics} 
The received signals of CUE and DUE as well as the echo signals of the radar have been defined in the previous subsection and in this subsection we first give the performance evaluation metric of the communication. The common stream rate for decoding at the $k$-th CUE can be expressed as
\begin{equation}
	R_{c,k}=\log_2\left(1+\frac{{{{\left| {\widetilde{\bf{h}}_k^H\left({{\rm vec}\left({\bf{W}}\right)}\circ {\bf a}_c\right)} \right|}^2}}}{{\sum\nolimits_{i = 1}^{K} {{{\left| {\widetilde{\bf{h}}_k^H\left({{\rm vec}\left({\bf{W}}\right)}\circ {\bf a}_i\right)} \right|}^2}}  + {\sigma _{{n_k}}^2}}}\right), \forall k.
\end{equation}
To ensure that CUEs can decode the common stream successfully, the following constraints must be satisfied
\begin{equation}
	\left({\bf p}_{k}\circ {\bf d}_1\right)^H{\bf c}_{k}\le R_{c,k},\forall k,
\end{equation}
where ${\bf c}_{k}=\left(C_{1,k},\cdots,C_{1,k},\cdots,C_{M,k},\cdots,C_{M,k}\right)^T\in\mathbb{C}^{MN\times1}$ denotes the $N$-fold of $\left(C_{1,k},\cdots,C_{M,k}\right)^T$ and $C_{m,k}$ denotes the equivalent common stream rate of the $m$-th DUE for the $k$-th CUE.

After the CUE decodes the common stream, it is split and encoded again and then sent to the DUE, the actual common stream rate at the $m$-th DUE can be expressed as
\begin{equation}
	C_{m} = \log_2\left(\frac{{{{\sum\limits_{m = 1}^{M}\!\left| \sum\limits_{k = 1}^{K}{\widetilde{\bf{g}}_{k,m}^H{\left({\rm vec}\left({\bf F}_{k}\right)\!\circ\!{\bf b}_m \!\circ\!{\bf p}_k\right)}} \right|}^2\!+\!{\sigma _{{n_m}}^2}}} }{\sum\limits_{j \ne m}^{M}\!{{{\left| \sum\limits_{k = 1}^{K}{\widetilde{\bf{g}}_{k,m}^H\left({\rm vec}\left({\bf F}_{k}\right)\!\circ\!{\bf b}_j \!\circ\!{\bf p}_k\right)} \right|}^2}}\!+\!{\sigma _{{n_m}}^2}}\right).
\end{equation}
Considering that the message $W_{d,m}$ is successfully transmitted over the BS-CUE link and the CUE-DUE link, the achievable rate of the $m$-th DUEs can be expressed as
\begin{equation}
	R_{d,m} = \min\left\{\sum\nolimits_{k = 1}^{K}\left({\bf p}_k\circ {\bf e}_m \right)^H{\bf c}_{k},C_{m}\right\}, \forall m,
\end{equation}
where ${\bf e}_m=\left(1,0,\cdots,0\right)\in \mathbb{C}^{NM\times 1}$ denotes the position index, taking the value $1$ at position $\left(k,1\right)$ and $0$ for the rest.

After CUEs decode the common stream, the common stream is removed from the received signal, followed by CUE decoding its own private stream. The private stream rate for the $k$-th CUEs can be expressed as
\begin{equation}
	R_{r,k}=\log_2\left(1+\frac{{{{\left| {\widetilde{\bf{h}}_k^H\left({{\rm vec}\left({\bf{W}}\right)}\circ {\bf a}_k\right)} \right|}^2}}}{{\sum\nolimits_{i \ne c,k}^{K} {{{\left| {\widetilde{\bf{h}}_k^H\left({{\rm vec}\left({\bf{W}}\right)}\circ {\bf a}_i\right)} \right|}^2}}  + {\sigma _{{n_k}}^2}}}\right), \forall k.
\end{equation}

For the performance evaluation metric of the sensing, we consider the radar mutual information (RMI)\cite{8579200,9832622}. The RMI represents the mutual information between the received radar signal and the channel parameters, which can be calculated as
\begin{equation}
	\begin{split}
		I_u &= I\left(y_u;{\bf z}_{k,u}|{\bf s}_d\right)\\
		&=\log_2\left(1+{\left|\sum\nolimits_{k=1}^{K}\sum\nolimits_{m=1}^{M}tp_{k,m,u}\right|^2}/{\sigma_u^2}\right),\forall u,
	\end{split}
\end{equation}
where $tp_{k,m,u}=\widetilde{\bf z}_{k,u}^H\left({\rm vec}\left({\bf F}_{k}\right) \circ {\bf b}_m\circ {\bf p}_k\right)$.

\subsection{Problem Formulation} 
In this paper, an optimization problem is constructed to maximize minimum RMI, and then design the optimization variables ${\cal S}=\left\{{\bf c}_{k}, {\bf W},{{\bf F}_{k}},{\bf p}_{k}\right\}$ by solving the following optimization problem
\begin{subequations}
	\begin{align}
		&\left( {{\textrm{P0}}} \right){\rm{:~}}{\mathop {\max~}\limits_{{\cal S}}\mathop {\min~}\limits_{U}}~I_{u}, \notag\\
		&~~{\textrm{s}}{\textrm{.t}}{\rm{.}}~~~\left\| {{\rm vec}\left({\bf W}\right)\circ{\bf c}_n} \right\|^2\leq P_t, \forall n,\label{BS}\\
		&~~~~~~~~~\left\| {{\rm vec}\left({\bf F}_k\right)\circ{\bf d}_n}\circ{\bf p}_k \right\|^2\leq P_t, \forall k,n,\label{pt}\\
		&~~~~~~~~~R_{r.k}\ge R_1,\forall k,\\
		&~~~~~~~~~R_{d,m}\ge R_2,\forall m,\\
		&~~~~~~~~~{\bf c}_{k}\succeq {\bf 0},\forall k,\\
		&~~~~~~~~~R_{c,k}\ge R_3,\forall k,\\
		&~~~~~~~~~\left({\bf p}_{k}\circ {\bf d}_1\right)^H{\bf c}_{k}\le R_{c,k},\forall k,\\
		&~~~~~~~~~\left({\bf 1}-{\bf p}_k\right)\circ\left({\rm vec}\left({\bf F}_k\right)\circ{\bf b}_m\right)=0,\forall k,m,\label{bs1}\\
		&~~~~~~~~~{\bf 0}\preceq {\bf p}_k\preceq {\bf 1},\forall k,\label{bs2}
	\end{align}
\end{subequations}
where constraints (\ref{BS}) and (\ref{pt}) denote the power constraints of the BS and CUEs, respectively, and the Eqs. (\ref{bs1}) and (\ref{bs2}) denote the user scheduling constraints\footnote{${\bf 0}$ and $\bf 1$ denote all-zero and all-one vectors or matrices of the same degree of latitude as their operational terms. (\ref{bs2}) is a relaxation constraint on ${\bf p}_k$, which we will recover in the following.}. $R_1$, $R_2$ and $R_3$ denote the rate thresholds for CUEs and DUE, respectively. It can be seen that problem (P0) is a nonconvex problem since coupling of variables and nonconvex operations. In next section, we will design a low-complexity distributed optimization algorithm based on the consensus ADMM.

\section{Distributed Design of Cooperative ISAC}
In this section, we transform and decompose the problem into several convex problems to solve, along with the design of low-complexity distributed algorithm, followed by giving a convergence analysis and a complexity analysis.

\subsection{Convex Transformation and Problem Reconstruction}
First, we handle the objective function, and since max-min is a non-convex operation, we introduce the auxiliary variable ${\gamma}$, and since $I_u$ is an increasing function, we transform the original problem into the following form
\begin{subequations}
	\begin{align}
		&\left( {{\textrm{P1}}} \right){\rm{:~}}{\mathop {\max~}\limits_{{\cal S},\gamma}}\gamma, \notag\\
		&~~~~~~~~~~{\textrm{s}}{\textrm{.t}}{\rm{.}}~~~{\rm(\ref{BS})-(\ref{bs2})},\\
		&~~~~~~~~~~~~~~~~~{\left|\sum\nolimits_{k=1}^{K}\sum\nolimits_{m=1}^{M}tp_{k,m,u}\right|^2}\ge\gamma,\forall u.
	\end{align}
\end{subequations}
Obviously problem (P1) remains a challenging nonconvex problem, which is solved below based on the consensus ADMM algorithmic framework\cite{6892987,7271111,huang2016consensus}. Specifically, introducing the set of auxiliary variables ${\cal D}=\left\{\eta_u, {{\bf{\Psi }}_k},{{\bf{\Gamma }}_n},{{\bf{T}}_{nk}},{{\bf{D}}_{mk}},{{\bf{V}}_{uk}},{{\bf{r}}_{kk}},{{\bf{f}}_{mk}},{{\bf{q}}_{kk}},{{\bf{y}}_{mk}},{{\bm{\varphi }}_{nk}},{{\bf{x}}_{uk}}\right\}$, problem (P1) can be further written as
\begin{subequations}
	\begin{align}
		&\left( {{\textrm{P1.1}}} \right){\rm{:~}}{\mathop {\max~}\limits_{{\cal S,D},\gamma}}\gamma, \notag\\
		&~~~~{\textrm{s}}{\textrm{.t}}{\rm{.}}~~~\left\| {{\rm vec}\left({{\bf{\Gamma }}_n}\right)\circ{\bf c}_n} \right\|^2\leq P_t, \forall n,\label{gao}\\
		&~~~~~~~~~~~\left\| {{\rm vec}\left({{\bf{T}}_{nk}}\right)\circ{\bf d}_n}\circ{{\bm{\varphi }}_{nk}} \right\|^2\leq P_t, \forall k,n,\label{pt}\\
		&~~~~~~~~~~~\widetilde R_{r.k}\ge R_1,\forall k,\\
		&~~~~~~~~~~~\sum\nolimits_{k = 1}^{K}\left({\bf y}_{mk}\circ {\bf e}_m \right)^H{\bf f}_{mk}\ge R_2,\forall m,\label{r1}\\
		&~~~~~~~~~~~\widetilde C_{m}\ge R_2,\forall m,\label{r2}\\
		&~~~~~~~~~~~{\bf c}_{k}\succeq {\bf 0},\forall k,\\
		&~~~~~~~~~~~\widetilde R_{c,k}\ge R_3,\forall k,\\
		&~~~~~~~~~~~\left({\bf q}_{kk}\circ {\bf d}_1\right)^H{\bf r}_{kk}\le \widetilde R_{c,k},\forall k,\\
		&~~~~~~~~~~~\left({\bf 1}-{\bf y}_{mk}\right)\circ\left({\rm vec}\left({\bf D}_{mk}\right)\circ{\bf b}_m\right)=0,\forall k,m,\\
		&~~~~~~~~~~~{\bf 0}\preceq {\bf q}_{kk}\preceq {\bf 1},\forall k,\\
		&~~~~~~~~~~~is_u\ge\eta_u,\forall u,\\
		&~~~~~~~~~~~{\eta_u}={\gamma},\forall u,\\
		&~~~~~~~~~~~{\bf W}={{\bf{\Psi }}_k},\forall k,\\
		&~~~~~~~~~~~{\bf W}={{\bf{\Gamma }}_n},\forall n,\\
		&~~~~~~~~~~~{\bf F}_k={{\bf{T}}_{nk}},\forall k,n,\\
		&~~~~~~~~~~~{\bf F}_k={{\bf{D}}_{mk}},\forall k,m,\\
		&~~~~~~~~~~~{\bf F}_k={{\bf{V}}_{uk}},\forall k,u,\\
		&~~~~~~~~~~~{\bf p}_k={{\bf{q}}_{kk}},\forall k,\\
		&~~~~~~~~~~~{\bf p}_k={{\bf{y}}_{mk}},\forall k,m,\\
		&~~~~~~~~~~~{\bf p}_k={{\bm{\varphi }}_{nk}},\forall k,n,\\
		&~~~~~~~~~~~{\bf p}_k={{\bf{x}}_{uk}},\forall k,u,\\
		&~~~~~~~~~~~{\bf c}_{k}={{\bf{r}}_{kk}},\forall k,\\
		&~~~~~~~~~~~{\bf c}_{k}={{\bf{f}}_{mk}},\forall k,m,
	\end{align}
\end{subequations}
where $\widetilde R_{r,k}$, $\widetilde R_{c,k}$, $\widetilde C_{m}$ and $is_u$ are given at the top of the next page\footnote{The constraints (\ref{r1}) and (\ref{r1}) are given by $R_{d,m}\ge R_2$. ${\bf d}_n\in\mathbb{C}^{NM\times1}$ denotes the position index, which takes the value $1$ at positions $(n,1),\cdots,(n,M)$ and $0$ for the rest and it is defined similarly to ${\bf c}_n$.}. With the introduction of auxiliary variables, problem (P1.1) is decomposed into several subproblems to be solved in the next subsection.
\begin{figure*}[ht] 
	\centering
	\begin{equation}
				\setlength{\abovedisplayskip}{3pt}
		\setlength{\belowdisplayskip}{3pt}
	\widetilde R_{r,k}=\log_2\left(1+\frac{{{{\left| {\widetilde{\bf{h}}_k^H\left({{\rm vec}\left({{\bf{\Psi }}_k}\right)}\circ {\bf a}_k\right)} \right|}^2}}}{{\sum\nolimits_{i \ne c,k}^{K} {{{\left| {\widetilde{\bf{h}}_k^H\left({{\rm vec}\left({{\bf{\Psi }}_k}\right)}\circ {\bf a}_i\right)} \right|}^2}}  + {\sigma _{{n_k}}^2}}}\right),\forall k,
	\end{equation}
		\centering
	\begin{equation}
				\setlength{\abovedisplayskip}{3pt}
		\setlength{\belowdisplayskip}{3pt}
\widetilde	R_{c,k}=\log_2\left(1+\frac{{{{\left| {\widetilde{\bf{h}}_k^H\left({{\rm vec}\left({{\bf{\Psi }}_k}\right)}\circ {\bf a}_c\right)} \right|}^2}}}{{\sum\nolimits_{i = 1}^{K} {{{\left| {\widetilde{\bf{h}}_k^H\left({{\rm vec}\left({{\bf{\Psi }}_k}\right)}\circ {\bf a}_i\right)} \right|}^2}}  + {\sigma _{{n_k}}^2}}}\right), \forall k,
	\end{equation}
		\centering
	\begin{equation}
				\setlength{\abovedisplayskip}{3pt}
		\setlength{\belowdisplayskip}{3pt}
		\widetilde C_{m}= \log_2\left(\frac{\sum\nolimits_{m=1}^M{{{\left| \sum\nolimits_{k = 1}^{K}{\widetilde{\bf{g}}_{k,m}^H{\left({\rm vec}\left({\bf D}_{mk}\right) \circ {\bf b}_m\circ {\bf y}_{mk}\right)}} \right|}^2+{\sigma _{{n_m}}^2}}} }{\sum\nolimits_{j \ne m}^{M}{{{\left| \sum\nolimits_{k = 1}^{K}{\widetilde{\bf{g}}_{k,m}^H\left({\rm vec}\left({\bf D}_{mk}\right) \circ {\bf b}_j\circ {\bf y}_{mk}\right)} \right|}^2}} +{\sigma _{{n_m}}^2}}\right),\forall m,
	\end{equation}
		\centering
\begin{equation}
			\setlength{\abovedisplayskip}{3pt}
	\setlength{\belowdisplayskip}{3pt}
 is_u={\left|\sum\nolimits_{k=1}^{K}\sum\nolimits_{m=1}^{M}\widetilde{\bf z}_{k,u}^H\left({\rm vec}\left({\bf V}_{uk}\right) \circ {\bf b}_m\circ {\bf x}_{uk}\right)\right|^2},\forall u.
\end{equation}
\hrulefill	
\end{figure*}
\subsection{Problem Solving and Algorithm Design}
In this subsection, due to the coupling of variables, we split the problem (P1.1) into several subproblems to solve and give the closed-form solutions in iterative form. 
\newcounter{name}
\setcounter{name}{0}
\stepcounter{name}
\newcounter{p}
\setcounter{p}{1}
\stepcounter{p}
\subsubsection{Solving for $\gamma$} Given the remaining variables and auxiliary variables, problem (P1.1) can be transformed into problem (P\thep), expressed as follows
	\begin{align}
		&\left( {{\textrm{P\thep}}} \right){\rm{:~}}{\mathop {\max~}\limits_{{\gamma}}}\gamma, \notag\\
		&~~~~~~~~~~{\textrm{s}}{\textrm{.t}}{\rm{.}}~~~\eta_u=\gamma,\forall u.
	\end{align}
Then, problem (P\thep) can be equivalently transformed into the following unconstrained problem (P\thep.1), which can be expressed as
	\begin{align}
		&\left( {{\textrm{P\thep.1}}} \right){\rm{:~}}{\mathop {\max~}\limits_{{\gamma,\xi_u}}}\gamma-\frac{\rho}{2}\sum\nolimits_{u = 1}^U\left(\eta_u-\gamma+\xi_u\right)^2,
	\end{align}
where $\rho$ denotes the penalty factor and $\xi_u$ denotes the error term between $\gamma$ and $\eta_u$. Defining the objective function of problem (P2.1) as $f\left(\gamma\right)$, it can be seen that it is a concave function with respect to the optimization variable $\gamma$. In order to determine the maximum value, we derive it and make it equal to zero
\begin{equation}
	{df\left(\gamma\right)}/{d\gamma}=-K\rho\gamma+1+\rho\sum\nolimits_{u = 1}^U\left(\eta_u+\xi_u\right)=0.
\end{equation}
Then, we get the optimal closed expression for $\gamma$ as follows
\begin{equation}
	\gamma=\frac{1+\rho\sum\nolimits_{u = 1}^U\left(\eta_u+\xi_u\right)}{K\rho}.
\end{equation}
\stepcounter{name}
\stepcounter{p}
\subsubsection{Solving for ${\bf W}$} Given the remaining variables and auxiliary variables, problem (P1.1) can be transformed into problem (P\thep), expressed as follows
\begin{subequations}
	\begin{align}
		\left( {{\textrm{P\thep}}} \right){\rm{:~}}{\mathop {{\rm find}}\limits_{{\bf W}}}~~&{\bf W},\notag\\
		{\textrm{s}}{\textrm{.t}}{\rm{.}}~~~&{\bf W}={{\bf{\Psi }}_k},\forall k,\\
		&{\bf W}={{\bf{\Gamma }}_n},\forall n.
	\end{align}
\end{subequations}
It can be further transformed into the unconstrained problem (P\thep.1) as follows
	\begin{align}
		\left( {{\textrm{P\thep.1}}} \right){\rm{:~}}{\mathop {{\min}~}\limits_{{\bf W},{{\bf{\Lambda }}_k},{{\bf{\Xi }}_n}}}f\left({\bf W}\right),
	\end{align}
where $f\left({\bf W}\right)\!=\!\sum\nolimits_{k = 1}^K\!\left\| {{{\bf{\Psi }}_k}\!-\!{\bf W}\!+\!{{\bf{\Lambda }}_k}} \right\|_F^2\!+\!\sum\nolimits_{n = 1}^N\!\left\| {{{\bf{\Gamma}}_n}\!-\!{\bf W}\!+\!{{\bf{\Xi }}_n}} \right\|_F^2$. ${{\bf{\Lambda }}_k}$ denotes the error term between ${\bf W}$ and ${{\bf{\Psi }}_k}$, and ${{\bf{\Xi }}_n}$ denotes the error term between ${\bf W}$ and ${{\bf{\Gamma}}_n}$. Since $f\left({\bf W}\right)$ is a convex function for ${\bf W}$, in order to obtain its minimum value, we take the partial derivative with respect to ${\bf W}$ and make it equal to zero
\begin{equation}
	\begin{split}
		{\partial f\left({\bf W}\right)}/{\partial{\bf W}}&=\sum\nolimits_{k = 1}^K\left( {{{\bf{\Psi }}_k}-{\bf W}+{{\bf{\Lambda }}_k}} \right)^*\\
		&+\sum\nolimits_{n = 1}^N\left( {{{\bf{\Gamma}}_n}-{\bf W}+{{\bf{\Xi }}_n}} \right)^*=0.
	\end{split}
\end{equation}
Then, we get the optimal closed expression for ${\bf W}$ as follows
\begin{equation}
	{\bf W}=\frac{\sum\nolimits_{k = 1}^K\left( {{{\bf{\Psi }}_k}+{{\bf{\Lambda }}_k}} \right)+\sum\nolimits_{n = 1}^N\left( {{{\bf{\Gamma}}_n}+{{\bf{\Xi }}_n}} \right)}{K+N}.	
\end{equation}
\stepcounter{name}
\stepcounter{p}
\subsubsection{Solving for ${\bf F}_k$} Given the remaining variables and auxiliary variables, problem (P1.1) can be transformed into problem (P\thep), expressed as follows
\begin{subequations}
	\begin{align}
		\left( {{\textrm{P\thep}}} \right){\textrm{:~}}{\mathop {{\rm find}}\limits_{{\bf F}_k}}~~&{\bf F}_k,\notag\\
		{\textrm{s}}{\textrm{.t}}{\textrm{.}}~~~&{\bf F}_k={{\bf{T}}_{nk}},\forall k,n,\\
		&{\bf F}_k={{\bf{V}}_{uk}},\forall k,u,\\
		&{\bf F}_k={{\bf{D}}_{mk}},\forall k,m.
	\end{align}
\end{subequations}
Similar to the process of solving for ${\bf W}$, the optimal closed expression for ${\bf F}_k$ can be expressed as
	\begin{align}
	&{\bf F}_k=\frac{1}{K+M+N}\left(\sum\nolimits_{n = 1}^N\left( {{{\bf{T }}_n}+{{\bf{O }}_n}} \right)\right.\\
	&~~~\left.+\sum\nolimits_{m = 1}^M\left( {{{\bf{D}}_{mk}}+{{\bf{B}}_{mk}}} \right)+\sum\nolimits_{u = 1}^U\left( {{{\bf{V}}_{uk}}+{{\bf{E}}_{uk}}} \right)\right),\forall k,\notag
	\end{align}
where ${{\bf{O}}_{nk}}$ denotes the error term between ${\bf F}_k$ and ${{\bf{T }}_{nk}}$, ${{\bf{B}}_{mk}}$ denotes the error term between ${\bf F}_k$ and ${{\bf{D }}_{mk}}$, and ${{\bf{E}}_{uk}}$ denotes the error term between ${\bf F}_k$ and ${{\bf{V }}_{uk}}$.
\stepcounter{name}
\stepcounter{p}

\subsubsection{Solving for ${\bf c}_k$} Given the remaining variables and auxiliary variables, problem (P1.1) can be transformed into problem (P\thep), expressed as follows
\begin{subequations}
	\begin{align}
		\left( {{\textrm{P\thep}}} \right){\textrm{:~}}{\mathop {{\rm find}}\limits_{{\bf c}_k}}~~&{\bf c}_k,\notag\\
		{\textrm{s}}{\textrm{.t}}{\rm{.}}~~~&{\bf c}_k={{\bf{f}}_{mk}},\forall k,m,\\
		&{\bf c}_k={{\bf{r }}_{kk}},\forall k.
	\end{align}
\end{subequations}
Similarly, the optimal closed expression for ${\bf c}_k$ can be expressed as
\begin{equation}
		{\bf c}_k=\frac{\sum\nolimits_{k = 1}^K\left( {{{\bf{r}}_{kk}}+{{\bf{m }}_{kk}}} \right)+\sum\nolimits_{m = 1}^M\left( {{{\bf{f}}_{mk}}+{{\bf{n}}_{mk}}} \right)}{K+M},\forall k,
\end{equation}
where ${{\bf{m}}_{kk}}$ denotes the error term between ${\bf c}_k$ and ${{\bf{r }}_{kk}}$, and ${{\bf{n}}_{mk}}$ denotes the error term between ${\bf c}_k$ and ${{\bf{f }}_{mk}}$.
\stepcounter{name}
\stepcounter{p}

\subsubsection{Solving for ${\bf p}_k$} Given the remaining variables and auxiliary variables, problem (P1.1) can be transformed into problem (P\thep), expressed as follows
\begin{subequations}
	\begin{align}
		\left( {{\textrm{P\thep}}} \right){\rm{:~}}{\mathop {{\rm find}}\limits_{{\bf p}_k}}~~&{\bf p}_k,\notag\\
		{\textrm{s}}{\textrm{.t}}{\rm{.}}~~~&{\bf p}_k={{\bf{q }}_{kk}},\forall k,\\
		&{\bf p}_k={{\bf{y}}_{mk}},\forall k,m,\\
		&{\bf p}_k={{\bm{\varphi}}_{nk}},\forall k,n,\\
		&{\bf p}_k={{\bf{x}}_{uk}},\forall k,u.
	\end{align}
\end{subequations}
Similarly, the optimal closed expression for ${\bf p}_k$\footnote{${\bf p}_k$ is a binary vector whose elements take integer values, and during the optimization process we relax it into continuous optimization variables, and at the end of the optimization we recover it by taking 1 when the elements take values greater than a threshold. The user scheduling variable recovery algorithm can be summarized as {\bf Algorithm \ref{alg1}}.} can be expressed as
	\begin{align}
		{\bf p}_k&=\frac{1}{K+M+N+U}\left(\sum\limits_{k = 1}^K\left( {{{\bf{q}}_{kk}}\!+\!{{\bf{l}}_{kk}}} \right)\!+\!\sum\limits_{m = 1}^M\left( {{{\bf{y}}_{mk}}\!+\!{{\bf{u}}_{mk}}} \right)\right.\notag\\
		&\left.+\sum\limits_{n = 1}^N\left( {{{\bm{\varphi}}_{nk}}\!+\!{{\bf\Pi}_{nk}}} \right)+\sum\limits_{u = 1}^U\left( {{{\bf{x}}_{uk}}\!+\!{{\bf{\Delta }}_{uk}}} \right)\right),\forall k,	
	\end{align}
where ${{\bf{l}}_{kk}}$ denotes the error term between ${\bf p}_k$ and ${{\bf{q }}_{kk}}$, ${{\bf{u}}_{mk}}$ denotes the error term between ${\bf p}_k$ and ${{\bf{y }}_{mk}}$, ${{\bf{\Pi}}_{nk}}$ denotes the error term between ${\bf p}_k$ and ${{\bm{\varphi }}_{nk}}$, and ${{\bf{\Delta}}_{uk}}$ denotes the error term between ${\bf p}_k$ and ${{\bf{x }}_{uk}}$.
\stepcounter{name}
\stepcounter{p}

\subsubsection{Solving for ${\eta_u}$} Given the remaining variables and auxiliary variables, problem (P1.1) can be transformed into problem (P\thep), expressed as follows
\begin{subequations}
	\begin{align}
		\left( {{\textrm{P\thep}}} \right){\rm{:~}}{\mathop {\rm find}\limits_{{\eta_u}}}~~&\eta_u, \notag\\
		{\textrm{s}}{\textrm{.t}}{\rm{.}}~~~&\eta_u=\gamma,\forall u,\\
		&is_u\ge\eta_u,\forall u.
	\end{align}
\end{subequations}
It can be further equated to the problem (P\thep.1) as follows
	\begin{align}
		&\left( {{\textrm{P\thep.1}}} \right){\rm{:~}}{\mathop {{\min}~}\limits_{{{\bf{\eta }}_u},{{{\xi }}_u}}}\left( {{{\bf{\eta}}_u}-{\gamma}+{{{\xi }}_u}} \right)^2,\notag\\
		&~~~~~~~~~~~~{\textrm{s}}{\textrm{.t}}{\rm{.}}~~~is_u\ge\eta_u,\forall u.
	\end{align}
The Lagrangian function of the problem can be expressed as
\begin{equation}
	\begin{split}
		{\cal L}_1\left({{\eta}_u}\right)&=\left( {{{\eta}_u}-{\gamma}+{{{\xi }}_u}} \right)^2+\lambda_u\left(\eta-is_u\right),\forall u,
	\end{split}
\end{equation}
where $\lambda_u$ denotes the Lagrange multiplier. Let $\partial {\cal L}_1\left({{\eta}_u}\right)/\partial \eta_u=0$, we have the following expression for the optimal closed solution of $\eta_u$
\begin{equation}
	\eta_u=\gamma-\xi_u-\lambda_u/2,\forall u,
\end{equation}
and the Lagrange multiplier is updated with the following rules
\begin{equation}
	\lambda_u^r=\left[\lambda_u^{r-1}-t_{u_1}\left(\eta_u-is_u\right)\right]^+,\forall u,
\end{equation}
where $t_{u_1}$ denotes the step size.
\stepcounter{name}
\stepcounter{p}

\subsubsection{Solving for ${{\bf {\Gamma}}_n}$} Given the remaining variables and auxiliary variables, problem (P1.1) can be transformed into problem (P\thep), expressed as follows
\begin{subequations}
	\begin{align}
		\left( {{\textrm{P\thep}}} \right){\rm{:~}}{\mathop {{\rm find}}\limits_{{{\bf{\Gamma }}_n}}}~~&{{\bf {\Gamma}}_n},\notag\\
		{\textrm{s}}{\textrm{.t}}{\rm{.}}~~~&{\textrm(\ref{gao})},\\
		&{\bf W}={{\bf{\Gamma }}_n},\forall k,n.
	\end{align}
\end{subequations}
It can be further equated to the problem (P\thep.1) as follows
	\begin{align}
		&\left( {{\textrm{P\thep.1}}} \right){\rm{:~}}{\mathop {{\min}~}\limits_{{{\bf{\Gamma }}_n},{{\bf{\Xi }}_n}}}\left\| {{{\bf{\Gamma}}_n}-{\bf W}+{{\bf{\Xi }}_n}} \right\|_F^2,\notag\\
		&~~~~~~~~~~~~{\textrm{s}}{\textrm{.t}}{\rm{.}}~~~{\textrm(\ref{gao})}.
	\end{align}
The Lagrangian function of the problem can be expressed as
\begin{equation}
	\begin{split}
		{\cal L}_2\left({{\bf{\Gamma }}_n}\right)\!=\!\left\| {{{\bf{\Gamma}}_n}\!-\!{\bf W}\!+\!{{\bf{\Xi }}_n}} \right\|_F^2\!+\!\theta_n\left(\left\|{\rm vec}\left({\bf\Gamma}_n\right)\!\circ\! {\bf c}_n\right\|^2\!-\!P_t\right),\forall n,
	\end{split}
\end{equation}
where $\theta_n$ denotes the Lagrange multiplier associated with constraint (\ref{gao}). It can be seen that function ${\cal L}_2\left({{\bf{\Gamma }}_n}\right)$ is convex with respect to variable ${{\bf{\Gamma }}_n}$. We derive its partial derivative with respect to ${{\bf{\Gamma }}_n}$ and make it equal to zero, expressed as
\begin{equation}
	\frac{\partial{\cal L}_2\left({{\bf{\Gamma }}_n}\right)}{\partial{{\bf{\Gamma }}_n}}=\left({{{\bf{\Gamma}}_n}-{\bf W}+{{\bf{\Xi }}_n}}\right)^*+\theta_n{{{\bf{\Gamma }}_n^*}}\circ {\bf C}_n=0,\forall n,
\end{equation}
where ${\bf C}_n\in\mathbb{C}^{N\times(K+1)}$ has a value of $1$ in the $n$-th row and $0$ in the rest. Then, we get
\begin{equation}
	\left({\bf 1}+\theta_n{\bf C}_n\right)\circ {{\bf{\Gamma }}_n}={\bf W}-{{\bf{\Xi }}_n},\forall n,
\end{equation}
To obtain ${{\bf{\Gamma }}_n}$, the following theorem is introduced.
\begin{theorem}
	Given matrices ${\bf A} \in \mathbb{C}^{N\times(K+1)}$ and ${\bf X} \in \mathbb{C}^{N\times(K+1)}$ it holds that ${\rm vec}\left({\bf A}\circ{\bf X}\right) = {\rm diag}({\bf A}){\rm vec}\left({\bf X}\right)$.
\end{theorem}
Then, according to {\bf Theorem 1}, we get the optimal closed expression for ${\bf \Gamma}_n$ as follows
\begin{equation}
			\setlength{\abovedisplayskip}{3pt}
	\setlength{\belowdisplayskip}{3pt}
	{\rm vec}\left({\bf \Gamma}_n\right)=\left({\rm diag}\left({\bf 1}+\theta_n{\bf C}_n\right)\right)^{-1}{\rm vec}\left({\bf W}-{{\bf{\Xi }}_n}\right),\forall n,
\end{equation}
where ${\rm diag}\left(\cdot\right)$ denotes the diagonalizing the matrix by columns, and the Lagrange multipliers are updated with the following rules
\begin{equation}
			\setlength{\abovedisplayskip}{3pt}
	\setlength{\belowdisplayskip}{3pt}
	\theta_n^{r}=\left[\theta_n^{r-1}-{t_{n_1}}\left(\left\|\mathrm{vec}\left(\mathbf{\Gamma}_n\right)\circ\mathbf{c}_n\right\|^2-P_t\right)\right]^+,\forall n,
\end{equation}
where $r$ denotes the the number of iterations and $t_{n_1}$ denotes the step size.
\stepcounter{name}
\stepcounter{p}

\subsubsection{Solving for ${{\bf {T}}_{nk}}$} Given the remaining variables and auxiliary variables, problem (P1.1) can be transformed into problem (P\thep), expressed as follows
\begin{subequations}
	\begin{align}
		\left( {{\textrm{P\thep}}} \right){\rm{:~}}{\mathop {{\rm find}}\limits_{{{\bf{T }}_{nk}}}}~~&{{\bf {T}}_{nk}},\notag\\
		{\textrm{s}}{\textrm{.t}}{\rm{.}}~~~&{\textrm(\ref{pt})},\\
		&{\bf T}_{nk}={{\bf{F}}_k},\forall k,n.\label{tnk}
	\end{align}
\end{subequations}
Similar to solving for ${\bf \Gamma}_n$, the optimal closed expression for ${\bf T}_{nk}$ can be expressed as
\begin{align}
			\setlength{\abovedisplayskip}{3pt}
	\setlength{\belowdisplayskip}{3pt}
		{\rm vec}\left({\bf T}_{nk}\right)\!=\!\left({\rm diag}\left({\bf 1}\!+\!\iota_{nk}\widetilde{\bf D}_{nk}\right)\right)^{-1}\!{\rm vec}\left({\bf F}_k\!-\!{{\bf{O }}_{nk}}\right),\forall k,n,
\end{align}
where $\widetilde{\bf D}_{nk}=\widetilde{\bf \Phi}_{nk}\circ{\bf D}_n$, $\widetilde{\bf \Phi}_{nk}$ denotes the reshaping of the vector ${\bm \varphi}_{nk}$ into a matrix of $N\times M$ by columns and ${\bf D}_n\in\mathbb{C}^{N\times M}$ has a value of $1$ in the $n$-th row and $0$ in the rest. The rules for updating Lagrange multipliers are as follows
\begin{align}
		\iota_{nk}^{r}\!=\!\left[\iota_{nk}^{r-1}\!-\!{t_{{nk}_1}}\left(\left\|\mathrm{vec}\left(\mathbf{T}_{nk}\right)\!\circ\!{\bm \varphi}_{nk}\!\circ\!\mathbf{d}_n\right\|^2\!-\!P_t\right)\right]^+,\forall k,n.
\end{align}
\stepcounter{name}
\stepcounter{p}
\subsubsection{Solving for ${{\bf {\Psi}}_k}$} Given the remaining variables and auxiliary variables, problem (P1.1) can be transformed into problem (P\thep), expressed as follows
\begin{subequations}
	\begin{align}
		\left( {{\textrm{P\thep}}} \right){\rm{:~}}{\mathop {{\rm find}}\limits_{{{\bf{\Psi }}_k}}}~~&{{\bf {\Psi}}_k},\notag\\
		{\textrm{s}}{\textrm{.t}}{\rm{.}}~~~&{\bf W}={{\bf{\Psi }}_k},\forall k,n,\\
		&\widetilde R_{r,k} \ge R_1,\forall k,\label{psi1}\\
		&\widetilde R_{c,k} \ge R_3,\forall k,\label{psi2}
	\end{align}
\end{subequations}
It can be further equated to the problem (P\thep.1) as follows
	\begin{align}
		&\left( {{\textrm{P\thep.1}}} \right){\rm{:~}}{\mathop {{\min}~}\limits_{{{\bf{\Psi }}_k},{{\bf{\Lambda }}_k}}}\left\| {{{\bf{\Psi}}_k}-{\bf W}+{{\bf{\Lambda }}_k}} \right\|_F^2,\notag\\
		&~~~~~~~~~~~~~~{\textrm{s}}{\textrm{.t}}{\rm{.}}~~~{\textrm(\ref{psi1})-(\ref{psi2})}.
	\end{align}
The Lagrangian function of the problem can be expressed as
\begin{align}
			\setlength{\abovedisplayskip}{3pt}
	\setlength{\belowdisplayskip}{3pt}
		&{\cal L}_3\left({{\bf{\Psi }}_k}\right)\!=\!\left\| {{{\bf{\Psi}}_k}-{\bf W}+{{\bf{\Lambda }}_k}} \right\|_F^2\!+\!\mu_k\mathbb{R}_1\sum\limits_{i\neq k}^K\!\left|\widetilde{{\bf h}}_k^H\left({\rm vec}\left({\bf \Psi}_k\right)\!\circ\!{\bf a}_i\right)\right|^2 \notag\\
		&-\mu_k\left|\widetilde{{\bf h}}_k^H\left({\rm vec}\left({\bf \Psi}_k\right)\circ{\bf a}_k\right)\right|^2\!+\!\pi_k\mathbb{R}_3\sum\limits_{i= k}^K\left|\widetilde{{\bf h}}_k^H\left({\rm vec}\left({\bf \Psi}_k\right)\circ{\bf a}_i\right)\right|^2\notag\\
		&-\pi_k\left|\widetilde{{\bf h}}_k^H\left({\rm vec}\left({\bf \Psi}_k\right)\circ{\bf a}_c\right)\right|^2+con_k,\forall k,
\end{align}
where $\mathbb{R}_1=2^{R_1-1}$, $\mathbb{R}_3=2^{R_3-1}$ and $con_k=(\mu_k\mathbb{R}_1+\pi_k\mathbb{R}_3){\sigma _{{n_k}}^2}$. It can be observed that ${\cal L}_3\left({{\bf{\Psi }}_k}\right)$ is nonconvex with respect to ${\rm vec}\left({\bf \Psi}_k\right)$. In the $r$th iteration, this subsection utilize the successive convex approximation (SCA) to obtain a lower bound for the nonconvex term of this Lagrangian function as
\begin{align}
			\setlength{\abovedisplayskip}{3pt}
	\setlength{\belowdisplayskip}{3pt}
	&\left(\left|\mathbf{\tilde{h}}_k^H\left(\mathrm{vec}\left(\mathbf{\Psi}_k\right)\circ\mathbf{a}_k\right)\right|^2\right)^{lb}=\left|\widetilde{\mathbf{h}}_k^H\left(\operatorname{vec}\left(\mathbf{\Psi}_k^{r}\right)\circ\mathbf{a}_k\right)\right|^2\label{lb1}\\
	&+\left(\left(\mathbf{H}_k^T\circ\mathbf{A}_k\right)\operatorname{vec}\left((\boldsymbol{\Psi}_k^r)^*\right)\right)^T\left(\operatorname{vec}\left(\boldsymbol{\Psi}_{k}\right)-\operatorname{vec}\left(\boldsymbol{\Psi}_{k}^{r}\right)\right),\forall k,\notag
\end{align}
and 
\begin{align}
			\setlength{\abovedisplayskip}{3pt}
	\setlength{\belowdisplayskip}{3pt}
	&\left(\left|\mathbf{\tilde{h}}_k^H\left(\mathrm{vec}\left(\mathbf{\Psi}_k\right)\circ\mathbf{a}_c\right)\right|^2\right)^{lb}=\left|\widetilde{\mathbf{h}}_k^H\left(\operatorname{vec}\left(\mathbf{\Psi}_k^{r}\right)\circ\mathbf{a}_c\right)\right|^2\label{lb2}\\
	&+\left(\left(\mathbf{H}_k^T\circ\mathbf{A}_c\right)\operatorname{vec}\left((\boldsymbol{\Psi}_k^r)^*\right)\right)^T\left(\operatorname{vec}\left(\boldsymbol{\Psi}_{k}\right)-\operatorname{vec}\left(\boldsymbol{\Psi}_{k}^{r}\right)\right),\forall k,\notag
\end{align}
where ${\bf H}_k=\widetilde{\bf h}_k\widetilde{\bf h}_k^H$ and ${\bf A}_k={\bf a}_k{\bf a}_k^H$. Replacing (\ref{lb1}) and (\ref{lb2}) in ${\cal L}_3\left({{\bf{\Psi }}_k}\right)$, we get its upper bound denoted as ${\cal L}_3\left({{\bf{\Psi }}_k}\right)^{ub}$. Then, we let ${\partial{\cal L}_3\left({{\bf{\Psi }}_k}\right)^{ub}}/{\partial{{\rm vec}\left({\bf{\Psi }}_k\right)}}=0$ and the optimal closed expression for ${\bf \Psi}_k$ can be expressed as
\begin{align}
{\rm vec}\left({\bf \Psi}_k\right)&=\left({\bf I}+{\bf H}_k\circ\widetilde{\bf A}_k\right)^{-1}\left({\rm vec}\left({\bf W}-{\bf \Lambda}_k\right)\right.\notag\\
&\left.+{\bf H}_k\circ\left(\mu_k{\bf A}_k+\pi_k{\bf A}_c\right){\rm vec}\left({\bf \Psi}_k^r\right)\right),\forall k,
\end{align}
where $\widetilde{\bf A}_k=\mu_k\mathbb{R}_1\sum\nolimits_{i\neq k}^K{\bf A}_i+\pi_k\mathbb{R}_3\sum\nolimits_{i = 1}^K{\bf A}_i$ and the Lagrange multipliers are updated with the following rules
\begin{align}
			\setlength{\abovedisplayskip}{3pt}
	\setlength{\belowdisplayskip}{3pt}
&\mu_k^r=\left[\mu_k^{r-1}-t_{k_{1}}\left(\mathbb{R}_1\left(\sum\nolimits_{i\neq k}^K\left|\mathbf{\tilde{h}}_k^H\left(\mathrm{vec}\left(\mathbf{\Psi}_k\right)\circ\mathbf{a}_k\right)\right|^2\!+\!\sigma_{n_k}^2\right)\right.\right.\notag\\
&\left.\left.-\left(\left|\mathbf{\tilde{h}}_k^H\left(\mathrm{vec}\left(\mathbf{\Psi}_k\right)\circ\mathbf{a}_k\right)\right|^2\right)^{lb}\right)\right]^+,\forall k,
\end{align}
and
\begin{align}
			\setlength{\abovedisplayskip}{3pt}
	\setlength{\belowdisplayskip}{3pt}
	&\pi_k^r=\left[\pi_k^{r-1}-t_{k_{2}}\left(\mathbb{R}_3\left(\sum\nolimits_{i= 1}^K\left|\mathbf{\tilde{h}}_k^H\left(\mathrm{vec}\left(\mathbf{\Psi}_k\right)\circ\mathbf{a}_k\right)\right|^2\!+\!\sigma_{n_k}^2\right)\right.\right.\notag\\
	&\left.\left.-\left(\left|\mathbf{\tilde{h}}_k^H\left(\mathrm{vec}\left(\mathbf{\Psi}_k\right)\circ\mathbf{a}_c\right)\right|^2\right)^{lb}\right)\right]^+,\forall k.
\end{align}
\stepcounter{name}
\stepcounter{p}
\subsubsection{Solving for ${{\bf {D}}_{mk}}$} Given the remaining variables and auxiliary variables, problem (P1.1) can be transformed into problem (P\thep), expressed as follows
\begin{subequations}
	\begin{align}
		&\left( {{\textrm{P\thep}}} \right){\rm{:~}}{\mathop {{\rm find}~}\limits_{{{\bf{D }}_{mk}}}}{{\bf {D}}_{mk}},\notag\\
		&~~~~{\textrm{s}}{\textrm{.t}}{\rm{.}}~~~{\bf F}_k={{\bf{D}}_{mk}},\forall k,m,\\
		&~~~~~~~~~~\left({\bf 1}-{\bf y}_{mk}\right)\circ\left({\rm vec}\left({\bf D}_{mk}\right)\circ{\bf b}_m\right)=0,\forall k,m,\\
		&~~~~~~~~~~~\widetilde C_{m}\ge R_2,\forall m.
	\end{align}
\end{subequations}
Similar to the process of solving for ${\bf \Psi}_k$, the optimal closed expression for ${\bf \Psi}_k$ can be expressed as
\begin{align}
	&{\rm vec}\left({\bf D}_{mk}\right)=\left({\bf I}+\tau_{mk}\mathbb{R}_2\sum\nolimits_{j\neq m}^M\left({\bf G}_{km}\circ\widetilde{\bf Y}_{jmk}\right)\right)^{-1}\notag\\
	&\left({\rm vec}\left({\bf F}_k-{\bf B}_{mk}\right)-\tau_{mk}\mathbb{R}_2\sum\nolimits_{j\neq m}^MJ_{ij}\widetilde{\bf g}_{k,m}^*\circ{\bf b}_j\circ{\bf y}_{mk}\right.\notag\\
	 &\left.+\tau_{mk}\left(J_{im}\widetilde{\bf g}_{k,m}^*\circ{\bf b}_m\circ{\bf y}_{mk}+\left({\bf G}_{km}\circ\widetilde{\bf Y}_{mk}\right){\rm vec}\left({\bf D}_{mk}^r\right)\right)\right.\notag\\
	 &\left.+{\bm\omega}_{mk}\circ\left({\bf y}_{mk}-{\bf 1}\right)\circ{\bf b}_m\right),\forall k,m,
\end{align}
where ${\bf G}_{km}=\widetilde{\bf g}_{k,m}\widetilde{\bf g}_{k,m}^H$, $\widetilde{\bf Y}_{jmk}=\left({\bf b}_j\circ{\bf y}_{mk}\right)\left({\bf b}_j\circ{\bf y}_{mk}\right)^H$, $J_{im}=\sum\nolimits_{i\neq k}^K\widetilde{\bf g}_{i,m}^H\left({\rm vec}\left({\bf D}_{mk}\right)\circ{\bf b}_m\circ{\bf y}_{mi}\right)$, and $\mathbb{R}_2=2^{R_2-1}$. The Lagrange multipliers are updated with the following rules
\begin{align}
	{\bm\omega}_{mk}^r\!=\!\left[{\bm\omega}_{mk}^{r-1}\!-\!t_{mk_1}\!\left({\bf 1}\!-\!{\bf y}_{mk}\right)\!\circ\!{\rm vec}\left({\bf D}_{mk}\right)\!\circ\!{\bf b}_m\right]^+,\forall k,m,
\end{align}
and
\begin{equation}
	\tau_{mk}^r=\left[\tau_{mk}^{r-1}-t_{mk_2}{cs}_{m}^{ub}\right]^+,\forall k,m,
\end{equation}
where ${cs}_{km}^{ub}=\mathbb{R}_2\sum\nolimits_{j \neq m}^M\!\left|\sum\nolimits_{k = 1}^K\!\widetilde{\bf{g}}_{k,m}^H{\left({\rm vec}\left({\bf D}_{mk}\right) \!\circ \!{\bf b}_j \!\circ\! {\bf y}_{mk}\right)}\right|^2\\\!+\!\sigma_{n_m}^2\!-\!\left(\left|\sum\nolimits_{k = 1}^K\!\widetilde{\bf{g}}_{k,m}^H\left({\rm vec}\left({\bf D}_{mk}\right) \!\circ\! {\bf b}_m \!\circ\! {\bf y}_{mk}\right)\right|^2\right)^{lb}$. The third term of ${cs}_{km}^{ub}$ can be expressed as
	\begin{align}
				\setlength{\abovedisplayskip}{3pt}
		\setlength{\belowdisplayskip}{3pt}
		&\left({cs}_{km}^{ub}\right)_{3_{rd}}=\left|\sum\nolimits_{k = 1}^K\widetilde{\bf{g}}_{k,m}^H\left({\rm vec}\left({\bf D}_{mk}^r\right)\circ {\bf b}_m\circ {\bf y}_{mk}\right)\right|^2\notag\\
		&+\left(J_{im}^*\widetilde{\bf g}_{k,m}\circ{\bf b}_m\circ{\bf y}_{mk}+\left({\bf G}_{km}^T\circ\widetilde{\bf Y}_{mk}\right){\rm vec}\left({\bf D}_{mk}^{*r}\right)\right)^H\notag\\
		&\left({\rm vec}\left({\bf D}_{mk}\right)-{\rm vec}\left({\bf D}_{mk}^r\right)\right),\forall k,m.
	\end{align}
\stepcounter{name}
\stepcounter{p}
		\setlength{\abovedisplayskip}{3pt}
\setlength{\belowdisplayskip}{3pt}
\subsubsection{Solving for ${{\bf {y}}_{mk}}$} Given the remaining variables and auxiliary variables, problem (P1.1) can be transformed into problem (P\thep), expressed as follows
\begin{subequations}
	\begin{align}
		&\left( {{\textrm{P\thep}}} \right){\rm{:~}}{\mathop {{\rm find}~}\limits_{{{\bf{y }}_{mk}}}}{{\bf {y}}_{mk}},\notag\\
		&~~~~{\textrm{s}}{\textrm{.t}}{\rm{.}}~~~{\bf p}_k={{\bf{y}}_{mk}},\forall k,m,\\
		&~~~~~~~~~~\left({\bf 1}-{\bf y}_{mk}\right)\circ\left({\rm vec}\left({\bf D}_{mk}\right)\circ{\bf b}_m\right)=0,\forall k,m,\\
		&~~~~~~~~~~\sum\nolimits_{k = 1}^{K}\left({\bf y}_{mk}\circ {\bf e}_m \right)^H{\bf f}_{mk}\ge R_2,\forall m,\\
		&~~~~~~~~~~~\widetilde C_{m}\ge R_2,\forall m.
	\end{align}
\end{subequations}
Similar to the process of solving for ${\bf D}_{mk}$, the optimal closed expression for ${\bf y}_{mk}$ can be expressed as
	\begin{align}
		&{\bf y}_{mk}=\left({\bf I}+o_{mk}\mathbb{R}_2\sum\nolimits_{j\neq m}^M\left({\bf G}_{km}\circ\widetilde{\bf D}_{jmk}^*\right)\right)^{-1}\notag\\
		&\left({\bf p}_k-{\bf u}_{mk}-o_{mk}\mathbb{R}_2\sum\nolimits_{j\neq m}^MJ_{ij}\widetilde{\bf g}_{k,m}^*\circ{\bf b}_j\circ{\rm vec}\left({\bf D}_{mk}^*\right)\right.\notag\\
		&\left.+o_{mk}\left(J_{im}\widetilde{\bf g}_{k,m}^*\circ{\bf b}_m\circ{\rm vec}\left({\bf D}_{mk}^*\right)\!+\!\left({\bf G}_{km}\circ\widetilde{\bf D}_{mk}^*\right){\bf y}_{mk}^r\right)\right.\notag\\
		&\left.+{\bm\varpi }_{mk}\circ{\rm vec}\left({\bf D}_{mk}^*\right)\circ{\bf b}_m-\sigma_{mk}\left({\bf f}_{mk}\circ{\bf e}_m\right)\right),\forall k,m,
	\end{align}
where $\widetilde{\bf D}_{jmk}^*=\left({\bf b}_j\circ{\rm vec}\left({\bf D}_{mk}^*\right)\right)\left({\bf b}_j\circ{\rm vec}\left({\bf D}_{mk}^*\right)\right)^H$ and the Lagrange multipliers are updated with the following rules
\begin{equation}
	\begin{split}
		\sigma_{mk}^r\!=\!\left[\sigma_{mk}^{r-1}\!-\!t_{mk_3}\!\left(R_2\!-\!\sum\nolimits_{k=1}^K\!\left({\bf y}_{mk}\!\circ\! {\bf e}_m \right)^H\!{\bf f}_{mk}\right)\right]^+,\!\forall k,m,
	\end{split}
\end{equation}
\begin{equation}
	{\bm\varpi}_{mk}^r=\left[{\bm\varpi}_{mk}^{r-1}-t_{mk_4}\left({\bf 1}-{\bf y}_{mk}\right)\!\circ\!{\rm vec}\left({\bf D}_{mk}\right)\!\circ\!{\bf b}_m\right]^+,\forall k,m,
\end{equation}
and
\begin{equation}
	o_{mk}^r=\left[o_{mk}^{r-1}-t_{mk_5}{co}_{m}^{ub}\right]^+,\forall k,m,
\end{equation}
where ${co}_{km}^{ub}=\mathbb{R}_2\sum\nolimits_{j \neq m}^M\!\left|\sum\nolimits_{k = 1}^K\widetilde{\bf{g}}_{k,m}^H{\left({\rm vec}\left({\bf D}_{mk}\right)\!\circ\!{\bf b}_j \!\circ\!{\bf y}_{mk}\right)}\right|^2\\\!+\!\sigma_{n_m}^2\!-\!\left(\left|\sum\nolimits_{k = 1}^K\!\widetilde{\bf{g}}_{k,m}^H\left({\rm vec}\left({\bf D}_{mk}\right)\!\circ\!{\bf b}_m \!\circ\!{\bf y}_{mk}\right)\right|^2\right)^{lb}$. The third term of ${co}_m^{ub}$ can be expressed as
	\begin{align}
		&\left({co}_{km}^{ub}\right)_{3_{rd}}=\left|\sum\nolimits_{k = 1}^K\widetilde{\bf{g}}_{k,m}^H\left({\rm vec}\left({\bf D}_{mk}\right) \circ {\bf b}_m\circ {\bf y}_{mk}^r\right)\right|^2\notag\\
		&+\tau_{mk}\left(J_{im}^*\widetilde{\bf g}_{k,m}\circ{\bf b}_m\circ{\rm vec}\left({\bf D}_{mk}\right)\!+\!\left({\bf G}_{km}^T\circ\widetilde{\bf D}_{mk}\right){\bf y}_{mk}^{r}\right)\notag\\
		&\left({\bf y}_{mk}-{\bf y}_{mk}^r\right),\forall k,m.
	\end{align}
\stepcounter{name}
\stepcounter{p}
		\setlength{\abovedisplayskip}{3pt}
\setlength{\belowdisplayskip}{3pt}
\subsubsection{Solving for ${{\bf {r}}_{kk}}$} Given the remaining variables and auxiliary variables, problem (P1.1) can be transformed into problem (P\thep), expressed as follows
\begin{subequations}
	\begin{align}
		\left( {{\textrm{P\thep}}} \right){\rm{:~}}{\mathop {{\rm find}}\limits_{{{\bf{r }}_{kk}}}}~~&{{\bf {r}}_{kk}},\notag\\
		{\textrm{s}}{\textrm{.t}}{\rm{.}}~~~&{\bf c}_k={{\bf{r }}_{kk}},\forall k,\\
		&\left({\bf q}_{kk}\circ{\bf d}_1\right)^H{\bf r}_{kk}\le R_{c,k},\forall k.
	\end{align}
\end{subequations}
Similar to the process of solving for ${\bf \Gamma}_n$, the optimal closed expression for ${\bf r}_{kk}$ can be expressed as
\begin{equation}
	{\bf r}_{kk}={\bf c}_k-{\bf m}_{kk}-\chi_{kk}\left({\bf q}_{kk}^*\circ{\bf d}_1\right),\forall k,
\end{equation}
and the Lagrange multipliers are updated with the following rules
\begin{equation}
	\chi_{kk}^r=\left[\chi_{kk}^{r-1}\!-\!t_{kk_1}\left(\left({\bf q}_{kk}\circ{\bf d}_1\right)^H{\bf r}_{kk}- R_{c,k}\right)\right]^+,\forall k.
\end{equation}
\stepcounter{name}
\stepcounter{p}
\subsubsection{Solving for ${{\bf {f}}_{mk}}$} Given the remaining variables and auxiliary variables, problem (P1.1) can be transformed into problem (P\thep), expressed as follows
\begin{subequations}
	\begin{align}
		\left( {{\textrm{P\thep}}} \right){\rm{:~}}{\mathop {{\rm find}}\limits_{{{\bf{f }}_{mk}}}}~~&{{\bf {f}}_{mk}},\notag\\
		{\textrm{s}}{\textrm{.t}}{\rm{.}}~~~&{\bf c}_k={{\bf{f }}_{mk}},\forall k,m,\\
		&\sum\nolimits_{k=1}^K\left({\bf y}_{mk}\circ {\bf e}_m \right)^H{\bf f}_{mk}\ge R_2,\forall m.
	\end{align}
\end{subequations}
Similar to the process of solving for ${\bf \Gamma}_n$, the optimal closed expression for ${\bf f}_{mk}$ can be expressed as
\begin{equation}
	{\bf f}_{mk}={\bf c}_k-{\bf n}_{mk}-\varphi_{mk}{\bf y}_{mk}^*\circ{\bf e}_m,\forall k,m,
\end{equation}
and the Lagrange multipliers are updated with the following rules
\begin{equation}
			\setlength{\abovedisplayskip}{3pt}
	\setlength{\belowdisplayskip}{3pt}
	\begin{split}
		\varphi_{mk}^r \!=\!\left[\varphi_{mk}^{r-1}\!-\!t_{kk_2}\!\left(R_2\!-\!\sum\nolimits_{k=1}^K\!\left({\bf y}_{mk}\!\circ\! {\bf e}_m \right)^H\!{\bf f}_{mk}\right)\right]^+,\!\forall k,m.
	\end{split}
\end{equation}
\stepcounter{name}
\stepcounter{p}
\subsubsection{Solving for ${{\bf {q}}_{kk}}$} Given the remaining variables and auxiliary variables, problem (P1.1) can be transformed into problem (P\thep), expressed as follows
\begin{subequations}
	\begin{align}
		&\left( {{\rm{P\thep}}} \right){\rm{:~}}{\mathop {{\rm find}~}\limits_{{{\bf{q }}_{kk}}}}{{\bf {q}}_{kk}},\notag\\
		&~~~~~~~~~~~{\rm{s}}{\rm{.t}}{\rm{.}}~~~{\bf p}_k={{\bf{q }}_{kk}},\forall k,\\
		&~~~~~~~~~~~~~~~~~\left({\bf q}_{kk}\circ{\bf d}_1\right)^H{\bf r}_{kk}\le R_{c,k},\forall k,\\
		&~~~~~~~~~~~~~~~~~~{\bf 0}\preceq {\bf q}_{kk}\preceq {\bf 1},\forall k.
	\end{align}
\end{subequations}
Similar to the process of solving for ${\bf \Gamma}_n$, the optimal closed expression for ${\bf q}_{kk}$ can be expressed as
\begin{equation}
	{\bf q}_{kk}={\bf p}_k-{\bf l}_{kk}-\nu_{kk}\left({\bf r}_{kk}^*\circ{\bf d}_1\right)+{\bm \zeta}_{kk}-{\bf \Omega}_{kk},\forall k,
\end{equation}
and the Lagrange multipliers are updated with the following rules
\begin{equation}
	\nu_{kk}^r=\left[\nu_{kk}^{r-1}-t_{kk_3}\left(\left({\bf q}_{kk}\circ{\bf d}_1\right)^H{\bf r}_{kk}- R_{c,k}\right)\right]^+,\forall k,
\end{equation}
\begin{equation}
	{\bm\zeta}_{kk}^r=\left[{\bm\zeta}_{kk}^{r-1}-t_{kk_4}{\bf q}_{kk}\right]^+,\forall k,
\end{equation}
and
\begin{equation}
	{\bm\Omega}_{kk}^r=\left[{\bm\Omega}_{kk}^{r-1}-t_{kk_5}\left({\bf q}_{kk}-{\bf 1}\right)\right]^+,\forall k.
\end{equation}
\stepcounter{name}
\stepcounter{p}
\subsubsection{Solving for ${{\bm {\varphi}}_{nk}}$} Given the remaining variables and auxiliary variables, problem (P1.1) can be transformed into problem (P\thep), expressed as follows
\begin{subequations}
	\begin{align}
		\left( {{\textrm{P\thep}}} \right){\rm{:~}}{\mathop {{\rm find}}\limits_{{{\bm{\varphi }}_{nk}}}}~~&{{\bm {\varphi}}_{nk}},\notag\\
		{\textrm{s}}{\textrm{.t}}{\rm{.}}~~~&{\textrm(\ref{pt})},\\
		&{\bf p}_k={{\bm{\varphi}}_{nk}},\forall k,n.
	\end{align}
\end{subequations}
Similar to the process of solving for ${\bf \Gamma}_n$, the optimal closed expression for ${\bm \varphi}_{nk}$ can be expressed as
\begin{equation}
	\begin{split}
			{\bm \varphi}_{nk}=\left({\rm diag}\left({\bf 1}+\delta_{nk}\widetilde{\bf {Td}}_{nk}\right)\right)^{-1}\left({\bf p}_k-{\bf\Pi }_{nk}\right),\forall k,n,
	\end{split}
\end{equation}
where $\widetilde{\bf {Td}}_{nk}={\rm vec}\left({\bf T}_{nk}^*\right)\circ{\rm vec}\left({\bf T}_{nk}\right)\circ{\bf d}_n$ and the Lagrange multipliers are updated with the following rules
\begin{align}
		\delta_{nk}^r \!=\!\left[\delta_{nk}^{r-1}\!-\! t_{{nk}_2}\left(\left\| {{\rm vec}\left({{\bf{T}}_{nk}}\right)\!\circ\!{\bf d}_n}\!\circ\!{{\bm{\varphi }}_{nk}} \right\|^2\!-\! P_t\right)\right]^+,\forall k,n.
\end{align}
\stepcounter{name}
\stepcounter{p}
\subsubsection{Solving for ${{\bf {x}}_{uk}}$} Given the remaining variables and auxiliary variables, problem (P1.1) can be transformed into problem (P\thep), expressed as follows
\begin{subequations}
	\begin{align}
		\left( {{\textrm{P\thep}}} \right){\rm{:~}}{\mathop {{\rm find}}\limits_{{{\bf{x }}_{uk}}}}~~&{{\bf {x}}_{uk}},\notag\\
		{\rm{s}}{\rm{.t}}{\rm{.}}~~~&{\bf p}_k={{\bf{x}}_{uk}},\forall k,u,\\
		&is_u\ge\eta_u,\forall u.
	\end{align}
\end{subequations}
It can be further equated to the problem (P\thep.1) as follows
\begin{subequations}
	\begin{align}
		&\left( {{\textrm{P\thep.1}}} \right){\rm{:~}}{\mathop {{\min}~}\limits_{{{\bf{x }}_{uk}},{{\bf{\Delta }}_{uk}}}}\left\| {{{\bf{x}}_{uk}}-{\bf p}_k+{{\bf{\Delta }}_{uk}}} \right\|_F^2,\notag\\
		&~~~~~~~~~~~~~~~{\textrm{s}}{\textrm{.t}}{\rm{.}}~~~is_u\ge\eta_u,\forall u.
	\end{align}
\end{subequations}
The Lagrangian function of the problem can be expressed as
\begin{align}
	{\cal L}_4\left({{\bf{x}}_{uk}}\right)\!=\!\left\| {{{\bf{x}}_{uk}}\!-\!{\bf p}_k\!+\!{{\bf{\Delta}}_{uk}}} \right\|_F^2\!+\!\vartheta_{uk}\left(\eta_u\!-\!is_u\right),\forall k,u,
\end{align}
Then, we let ${\partial{\cal L}_4\left({{\bf{x }}_{uk}}\right)}/{\partial{{\bf{x }}_{uk}}}=0$ and the optimal closed expression for ${\bf x}_{uk}$ can be expressed as
\begin{equation}
	{\bf x}_{uk} = {\bf p}_k-{\bm \Delta}_{uk}+\vartheta_{uk}\left(\hat{\bf z}_{uk}^*\hat a_{uk}+\hat{\bf Z}_{uk}^*{\bf x}_{uk}^r\right),\forall k,u,
\end{equation}
where $\hat{\bf z}_{uk}=\sum\nolimits_{m=1}^M\widetilde{\bf z}_{k,u}^H\circ\left({\rm vec}\left({\bf V}_{uk}\right)\circ{\bf b}_m\right)^H$, $\hat{\bf Z}_{uk}=\hat{\bf z}_{uk}\hat{\bf z}_{uk}^H$ and $\hat a_{uk}=\sum\nolimits_{i\neq k}^K\sum\nolimits_{m=1}^M\widetilde{\bf z}_{i,u}^H\left({\rm vec}\left({\bf V}_{ui}\right) \circ {\bf b}_m\circ {\bf x}_{ui}\right)$. The Lagrange multipliers are updated with the following rules
\begin{align}
	&\vartheta_{uk}^r=\left[\vartheta_{uk}^{r-1}-t_{uk_{1}}\left(\eta_u-is_u\right)\right]^+,\forall k,u.
\end{align}
\stepcounter{name}
\stepcounter{p}
\subsubsection{Solving for ${{\bf {V}}_{uk}}$} Given the remaining variables and auxiliary variables, problem (P1.1) can be transformed into problem (P\thep), expressed as follows
\begin{subequations}
	\begin{align}
		\left( {{\textrm{P\thep}}} \right){\rm{:~}}{\mathop {{\rm find}}\limits_{{{\bf{V }}_{uk}}}}~~&{{\bf {V}}_{uk}},\notag\\
		{\textrm{s}}{\textrm{.t}}{\rm{.}}~~~&{\bf F}_k={{\bf{V }}_{uk}},\forall k,u,\\
		&is_u\ge\eta_u,\forall u.
	\end{align}
\end{subequations}
Similar to the process of solving for ${\bf x}_{uk}$, the optimal closed expression for ${\bf V}_{uk}$ can be expressed as
\begin{align}
	{\rm vec}\left({\bf V}_{uk}\right)={\rm vec}\left({\bf F}_{k}-{\bf E}_{uk}\right)+\kappa_{uk}\widetilde{\bf S}_{uk},\forall k,u,
\end{align}
where $\widetilde{\bf S}_{uk}={\bf s}_{uk}^*e_{uk}+{\bf S}_{uk}^*{\rm vec}\left({\bf V}_{uk}^r\right)$, ${\bf s}_{uk}=\sum\nolimits_{m=1}^M\widetilde{\bf z}_{k,u}\circ{\bf x}_{uk}\circ{\bf b}_m\in\mathbb{C}^{NM\times1}$, ${\bf S}_{uk}={\bf s}_{uk}{\bf s}_{uk}^H$ and $e_{uk}=\sum\nolimits_{i\neq k}^K\sum\nolimits_{m=1}^M\widetilde{\bf z}_{i,u}^H\left({\rm vec}\left({\bf V}_{ui}\right) \circ {\bf b}_m\circ {\bf x}_{ui}\right)$. The Lagrange multipliers are updated with the following rules
\begin{align}
		&\kappa_{uk}^r=\left[\kappa_{uk}^{r-1}-t_{uk_{2}}\left(\eta_u-is_u\right)\right]^+,\forall k,u.
\end{align}
\stepcounter{name}
\stepcounter{p}
\subsubsection{Error terms update} The rules for updating the error terms mentioned above are as follows

{\noindent The error term update rule of $\eta_u$}
\begin{equation}
	\xi_u \leftarrow \xi_u+\eta_u-\gamma,\forall u.\label{e1}
\end{equation}
The error term update rule of $\bf W$
\begin{equation}
	\left\{\begin{aligned}
		{{\bf{\Lambda }}_k}\leftarrow{{\bf{\Lambda }}_k}+{{\bf{\Psi }}_k}-{\bf W},\forall k,\\
		{{\bf{\Xi }}_n}\leftarrow{{\bf{\Xi }}_n}+{{\bf{\Gamma }}_n}-{\bf W},\forall n.
	\end{aligned}\right.
\end{equation}
The error term update rule of ${\bf F}_k$
\begin{equation}
	\left\{\begin{aligned}
		&{{\bf{O}}_{nk}}\leftarrow{{\bf{O}}_{nk}}+{{\bf{T }}_{nk}}-{\bf F}_k,\forall k,n,\\
		&{{\bf{B}}_{mk}}\leftarrow{{\bf{B}}_{mk}}+{{\bf{D}}_{mk}}-{\bf F}_k,\forall k,m,\\
		&{{\bf{E}}_{uk}}\leftarrow{{\bf{E}}_{uk}}+{{\bf{V}}_{uk}}-{\bf F}_k,\forall k,u.
	\end{aligned}
	\right.
\end{equation}
The error term update rule of ${\bf c}_k$
\begin{equation}
	\left\{\begin{aligned}
		&{{\bf{m}}_{kk}}\leftarrow{{\bf{m}}_{kk}}+{{\bf{r}}_{kk}}-{\bf c}_k,\forall k,\\
		&{{\bf{n}}_{mk}}\leftarrow{{\bf{n}}_{mk}}+{{\bf{f}}_{mk}}-{\bf c}_k,\forall k,m.
	\end{aligned}\right.
\end{equation}
The error term update rule of ${\bf p}_k$
\begin{equation}
	\left\{\begin{aligned}
		&{{\bf{l}}_{kk}}\leftarrow{{\bf{l}}_{kk}}+{{\bf{q}}_{kk}}-{\bf p}_k,\forall k,\\
		&{{\bf{u}}_{mk}}\leftarrow{{\bf{u}}_{mk}}+{{\bf{y}}_{mk}}-{\bf p}_k,\forall k,m,\\
		&{{\bf\Pi}_{nk}}\leftarrow{{\bf\Pi}_{nk}}+{{\bm{\varphi}}_{mk}}-{\bf p}_k,\forall k,n,\\
		&{{\bm{\Delta}}_{uk}}\leftarrow{{\bm{\Delta}}_{uk}}+{{\bm{x}}_{uk}}-{\bf p}_k,\forall k,u.
	\end{aligned}\right.\label{en}
\end{equation}
Based on the above subproblems, {\bf Algorithm \ref{alg2}} is proposed in this section, which generalizes the distributed cooperative ISAC algorithm under the transmissive RIS transceiver architecture. Specifically, the optimization variables are determined by solving each subproblem independently while keeping the values of the other variables fixed, and then updating the error terms accordingly\footnote{Note that the services of CUEs for DUEs are cooperative, and CUEs cooperate with each other by interacting with the BS, while the optimization variables are solved at CUEs in a distributed manner, which reduces the complexity of the algorithms and improves the performance of the system.}. The subproblems are then solved by iteration until convergence is achieved for the entire problem.

\begin{algorithm}[htbp]
	\caption{User Scheduling Variable Recovery Algorithm}
	\label{alg1}
	\begin{algorithmic}[1]
		\STATE {\bf{Input}}:  $N$-fold expansion variable ${\bf p}_k$ and threshold $th$.
		\STATE {\bf{for}} $k=1,2,\cdots,K$\\
	~~~{\bf{for}} $m=1,2,\cdots,M$\\
	~~~~~~{\bf if} $sum({\bf p}_k(1+(m-1)N:mN,k))>=Nth$\\
	~~~~~~~~~${\bf p}_k(1+(m-1)N:mN,k) = ones(N,1)$; \\
	~~~~~~{\bf else}\\
	~~~~~~~~~${\bf p}_k(1+(m-1)N:mN,k) = zeros(N,1)$; \\
	~~~~~~{\bf end}\\
	~~~~~~~~~${\bm \rho}_k(m,k)={\bf p}_k(mN,k)$;\\
	~~~{\bf end}\\
	{\bf end}
		\STATE {\bf{Output}} The user scheduling variable ${\bm\rho}_k$.
	\end{algorithmic}  
\end{algorithm}

\begin{algorithm}[htbp]
	\caption{Distributed Cooperative ISAC Algorithm}
	\label{alg2}
	\begin{algorithmic}[1]
		\STATE {\bf{Initialization}}: ${\cal S}^{0}$, ${\cal D}^{0}$, $\gamma^{0}$, convergence threshold $\varepsilon$ and iteration index $r = 0$.
		\REPEAT
		\STATE To obtain ${\cal S}^{r}$ and $\gamma^{r}$, utilize the derived closed-form solutions by solving problems (P2)-(P6).

		\STATE To obtain ${\cal D}^{r}$, utilize the derived closed-form solutions and the Lagrange multipliers update rules by solving problems (P7)-(P18).
		\STATE Recovering user scheduling variables utilizing {\bf Algorithm \ref{alg1}}.
		\STATE Update the error terms by Eq.  (\ref{e1})-(\ref{en}).
		\STATE $r \leftarrow r + 1$.
		\UNTIL The fractional decrease of the objective value is
		below a threshold $\varepsilon$.
		\STATE {\bf{return}} The distributed cooperative ISAC design solution.
	\end{algorithmic}  
\end{algorithm}

\subsection{Convergence \& Computational Complexity Analysis}
{\it 1) Computational Complexity Analysis:} The overall computational complexity of {\bf Algorithm \ref{alg2}} is ${\cal O}\left( {\log \left( {1/\varepsilon } \right)} \left(3MK+2NK+4KK+5K+N+2\right)\right)$, where $\varepsilon$ denotes the precision of stopping the iteration.

{\it 2) Convergence Analysis:} The convergence analysis of {\bf Algorithm \ref{alg2}} can be proved by {\bf Theorem 2}. 
\begin{theorem}
Denote ${\bf x}^r$ and ${\bf z}_i^r$ the updates obtained at the r-th iteration of {\bf Algorithm \ref{alg2}}. Assume that the ${\bf z}_i^r$ is well-defined for all $r$ and $i$, and that
		\setlength{\abovedisplayskip}{3pt}
\setlength{\belowdisplayskip}{3pt}
\[\lim_{r\to+\infty}(\boldsymbol{z}_i^r-\boldsymbol{x}^r)=0,\forall i=1,...,{\cal I},\]
and
\[\lim_{r\to+\infty}(\boldsymbol{x}^{r+1}-\boldsymbol{x}^r)=0,\]
then any limit point of $\left\{{\bf x}^r\right\}$ is a Karush-Kuhn-Tucker point of the initial optimization problem{\rm\cite{huang2016consensus}}.
\end{theorem}
\section{Numerical Results}
In this section, numerical simulations of the proposed algorithm are performed to demonstrate its effectiveness. A three-dimensional polar coordinate system is used, the BS is located at (0{\rm{m}}, 0{\rm{m}}, 50{\rm{m}}), $K=3$ CUEs are distributed at coordinates (50\rm{m}, 50\rm{m}, 10\rm{m}), (-50\rm{m}, 50\rm{m}, 10\rm{m}), and (0\rm{m}, -50$\sqrt{\rm 2}$\rm{m}, 10\rm{m}), respectively, and $M=5$ DUEs are distributed far from the BS, with coordinates of (150\rm{m}, 50\rm{m}, 5\rm{m}), (25\rm{m}, 200\rm{m}, 5\rm{m}), (-25\rm{m}, 200\rm{m}, 5\rm{m}), (-150\rm{m}, -50\rm{m}, 5\rm{m}), and (0\rm{m}, -150$\sqrt{\rm 2}$\rm{m}, 5\rm{m}), respectively. The carrier frequency is 3 GHz, the system bandwidth is 20MHz, the noise power density is -90 dBm and the convergence precision is $10^{ - 3}$. 

In this paper, we compare the performance of the proposed algorithm and othe benchmarks as follows: (1) {\it{Traditional Cooperative Transceiver with ADMM}} (TCADMM): this scheme uses traditional multi-antenna transceivers for CUEs and the network operates in a cooperative manner with a power constraint that can be expressed as $\sum\nolimits_{i = 1}^{\cal \hat K}{\rm tr}\left({\bf W}_i\right)\le NP_{t}$. (2) {\it{Independent ADMM}} (IADMM): this scenario network operates in an independent manner, i.e., CUEs do not cooperate with each other. (3) {\it{Cooperative Semidefinite Programming}} (CSDP): this scheme utilizes the SDP algorithm to solve the cooperative ISAC problem. (4) {\it{Traditional Cooperative Transceiver with SDP}} (TCSDP): this scheme utilizes SDP algorithm and the traditional transceivers for CUEs to solve the cooperative ISAC problem. (5) {\it{Independent SDP}} (ISDP): this scheme utilizes SDP algorithm and the network operates in a independent manner.

\begin{figure}[!htpb]
	\centerline{\includegraphics[width=5.5cm]{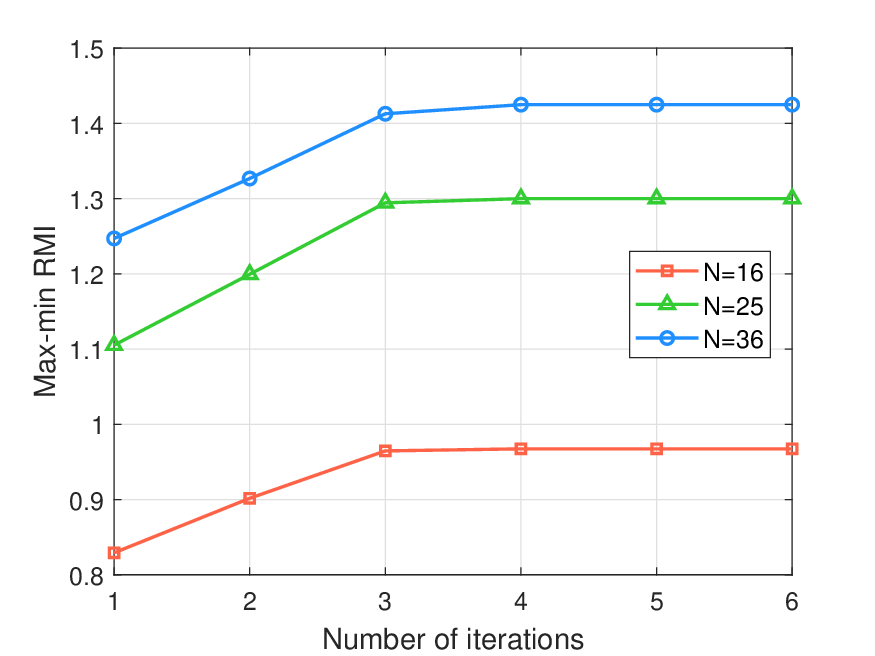}}
	\caption{Convergence process: Max-min RMI versus number of iterations under different TRIS elements ($P_t = 1 {\rm W}$).}\label{conv}
\end{figure}
Firstly, the convergence performance of the proposed algorithm is illustrated in Fig. \ref{conv}, where it can be seen that a fast convergence effect can be achieved with the number of iterations up to 4. Furthermore, the effect of different TRIS elements on the system performance is evaluated, and the results show that more elements lead to an improvement in the perceptual performance I, which is attributed to the spatial diversity gain of more TRIS elements.

\begin{figure}[!htbp]
	\centering
	\begin{minipage}[t]{1\linewidth} 
		\centering
		\begin{tabular}{@{\extracolsep{\fill}}c@{}c@{}@{\extracolsep{\fill}}}
			\includegraphics[width=0.52\linewidth]{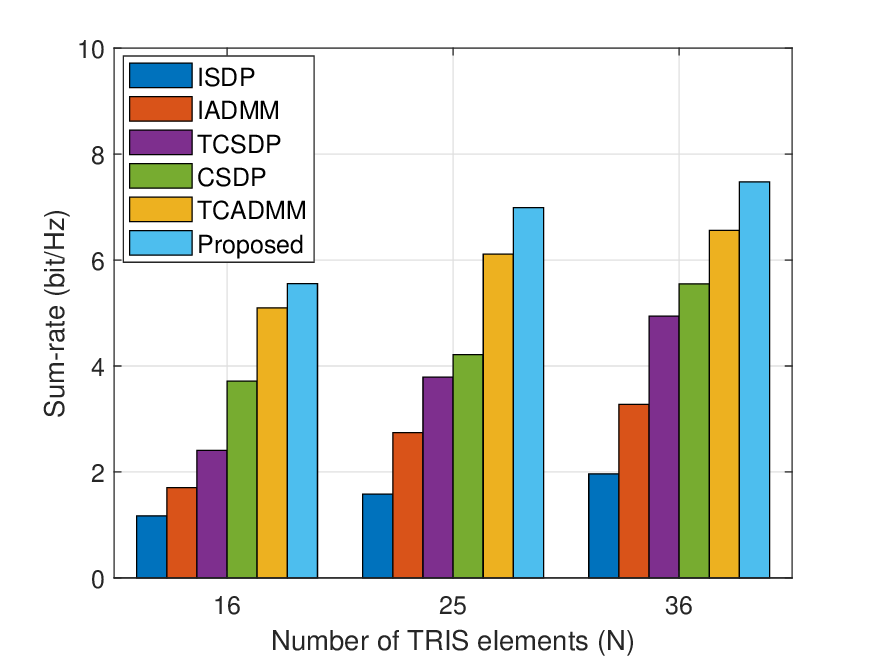}&
			\includegraphics[width=0.52\linewidth]{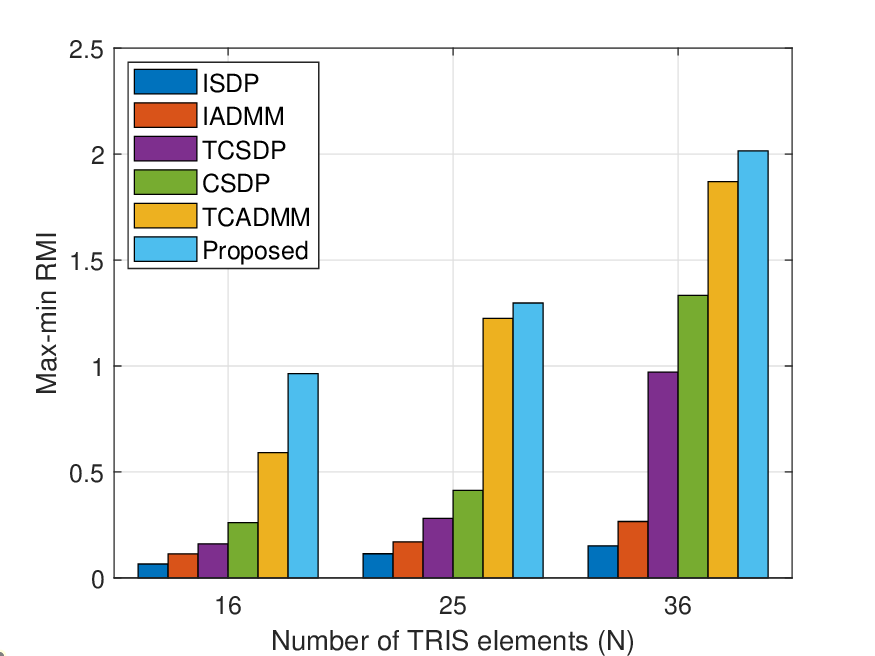}\\
			(a) Sum-rate vs. $N$ & (b) Max-min RMI vs. $N$\\
		\end{tabular}
	\end{minipage}
	\caption{C\&S performance under different TRIS elements ($P_{t}=1{\rm W}$).}\label{per}
\end{figure}
Then, the trend of the C\&S performance with the number of TRIS elements is exhibited in Fig. \ref{per}. It can be seen that as the number of TRIS elements increases, the C\&S performance of the proposed scheme improves and outperforms the other schemes, which is due to the fact that more TRIS elements bring higher diversity gain, and the proposed scheme adopts TRIS transceivers that can bring higher aperture efficiency to realize higher sum-rate and RMI. Due to the characteristics of the TRIS transceiver's modulation method, multiple streams of communication can be realized, which lead to better performance than the traditional transceiver.

\begin{figure}[!htbp]
	\centering
	\begin{minipage}[t]{1\linewidth} 
		\centering
		\begin{tabular}{@{\extracolsep{\fill}}c@{}c@{}@{\extracolsep{\fill}}}
			\includegraphics[width=0.52\linewidth]{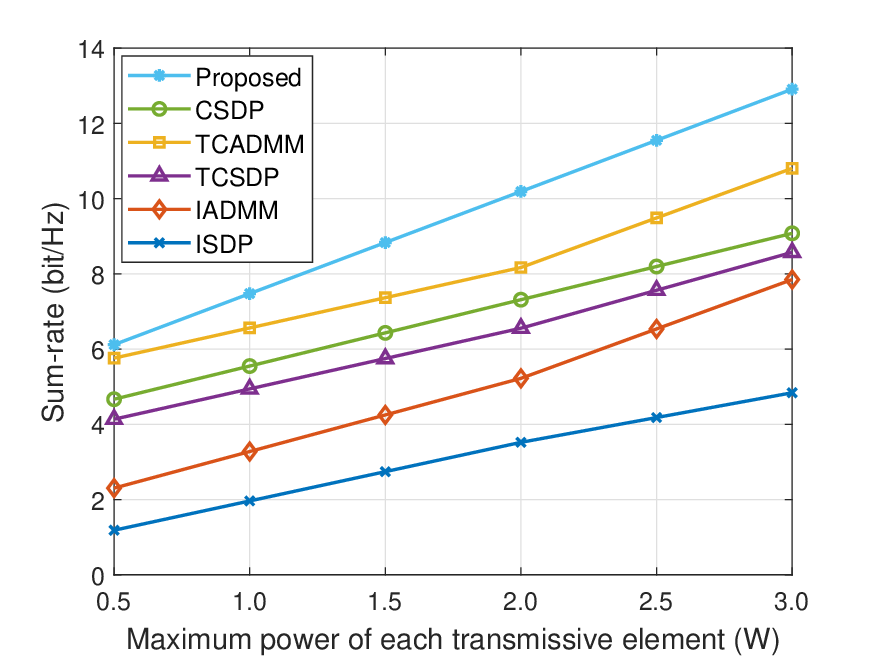}&
			\includegraphics[width=0.52\linewidth]{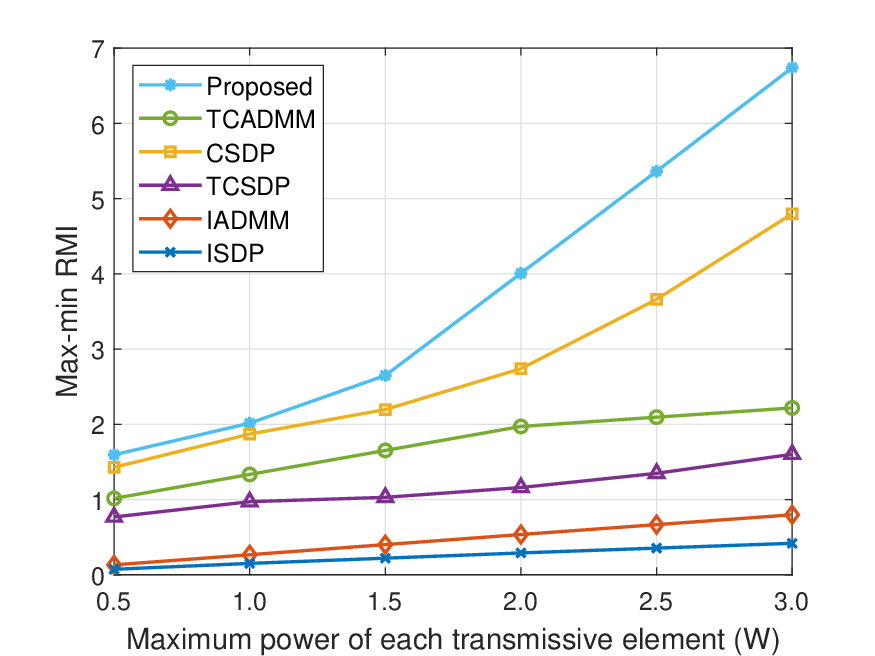}\\
			(a) Sum-rate vs. $P_t$ & (b) Max-min RMI vs. $P_t$\\
		\end{tabular}
	\end{minipage}
	\caption{C\&S performance varies with the maximum power of each transmissive element ($N=36$).}\label{per2}
\end{figure}

Subsequently, the relationship between the maximum available power of each TRIS element and the C\&S performance is analyzed. In Fig. \ref{per2}, it can be seen that as the maximum available transmit power of the TRIS element increases, the sum-rate and RMI show an increasing trend, which is due to the fact that a larger transmit power enables the system to provide a stronger gain to the user. Compared to the traditional transceiver scheme (TCADMM), the proposed algorithm has a communication performance gain of 19.47\% and a sensing gain of 40.48\%. Furthermore, system performance gains are higher in cooperative schemes than in independent schemes.

\begin{figure}[!htpb]
	\centerline{\includegraphics[width=5.5cm]{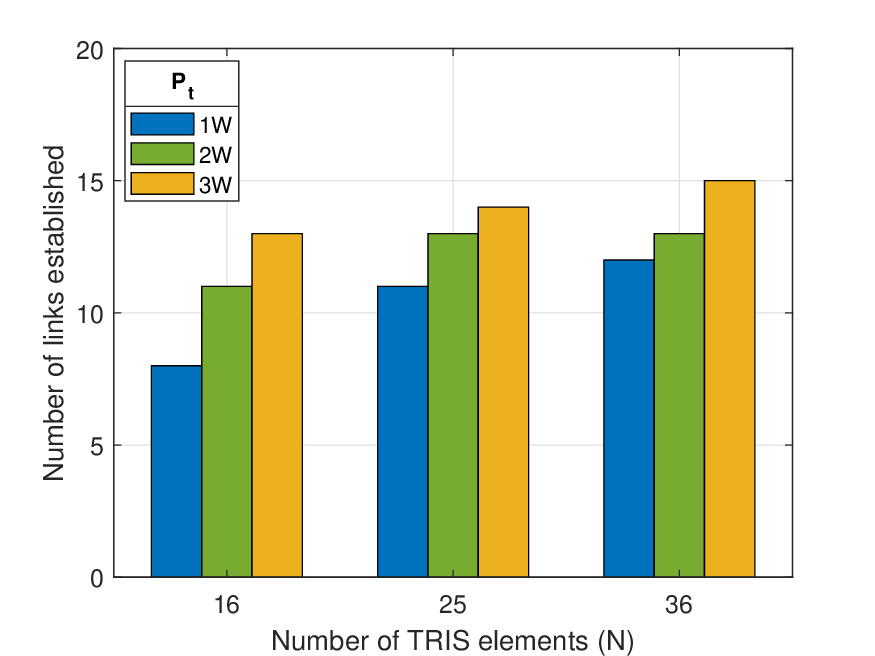}}
	\caption{The number of links established varies with the number of TRIS elements and the maximum power of each transmissive element.}\label{lin}
\end{figure}
\begin{figure}
	\centering
	\begin{minipage}[t]{1\linewidth} 
		\centering
		\begin{tabular}{@{\extracolsep{\fill}}c@{}c@{}@{\extracolsep{\fill}}}
			\includegraphics[width=0.52\linewidth]{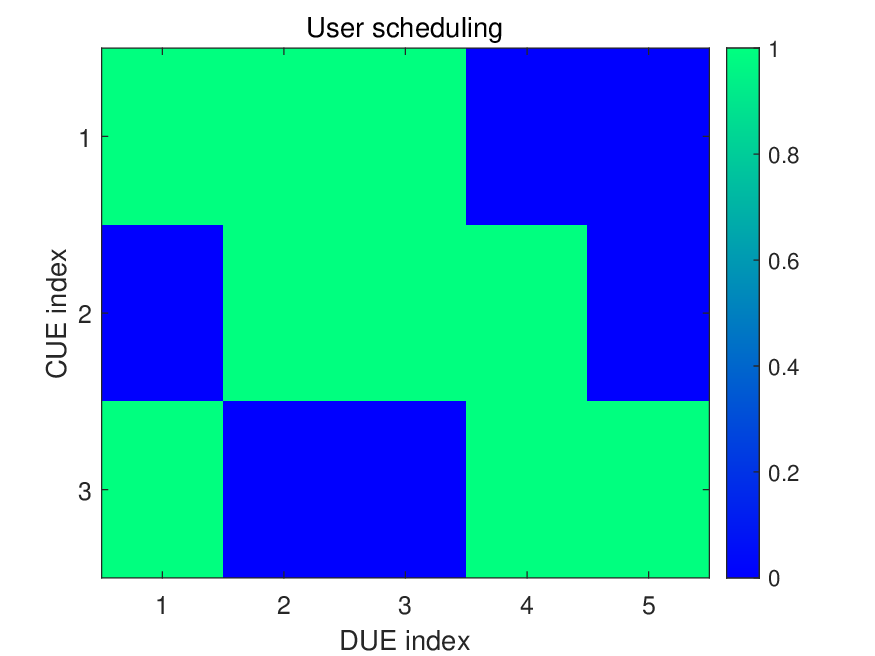}&
			\includegraphics[width=0.52\linewidth]{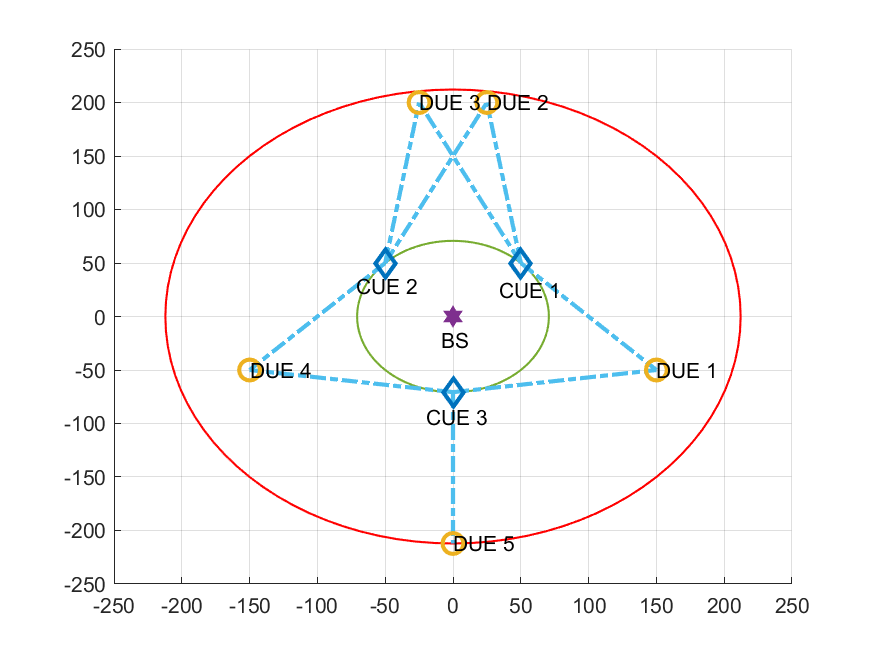}\\
			(a) Sum-rate vs. $P_t$ & (b) Max-min RMI vs. $P_t$\\
		\end{tabular}
	\end{minipage}
	\caption{Links established results ($P_t=1 W, N=16$).}\label{con}
\end{figure}
In order to explore the impact factors of cooperative CUEs on network coverage, we elaborate in Fig. \ref{lin}. It can be seen that the number of established links of DUEs tends to increase with the increase of the maximum transmit power of TRIS elements and elements, which provides a design guideline to enhance the network coverage. Moreover, the links established condition between the CUEs and the DUEs is visualized in Fig. \ref{con}. The results show that DUEs that are closer to the CUEs are prioritized to establish links, which is consistent with the practical scenario.

\begin{figure}[!htbp]
	\centering
	\begin{minipage}[t]{1\linewidth} 
		\centering
		\begin{tabular}{@{\extracolsep{\fill}}c@{}c@{}@{\extracolsep{\fill}}}
			\includegraphics[width=0.52\linewidth]{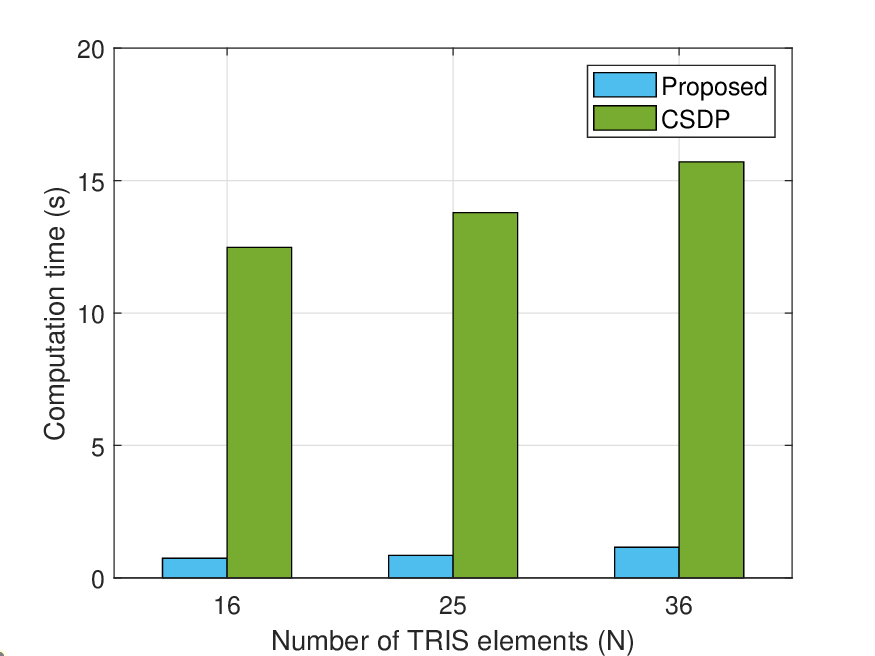}&
			\includegraphics[width=0.52\linewidth]{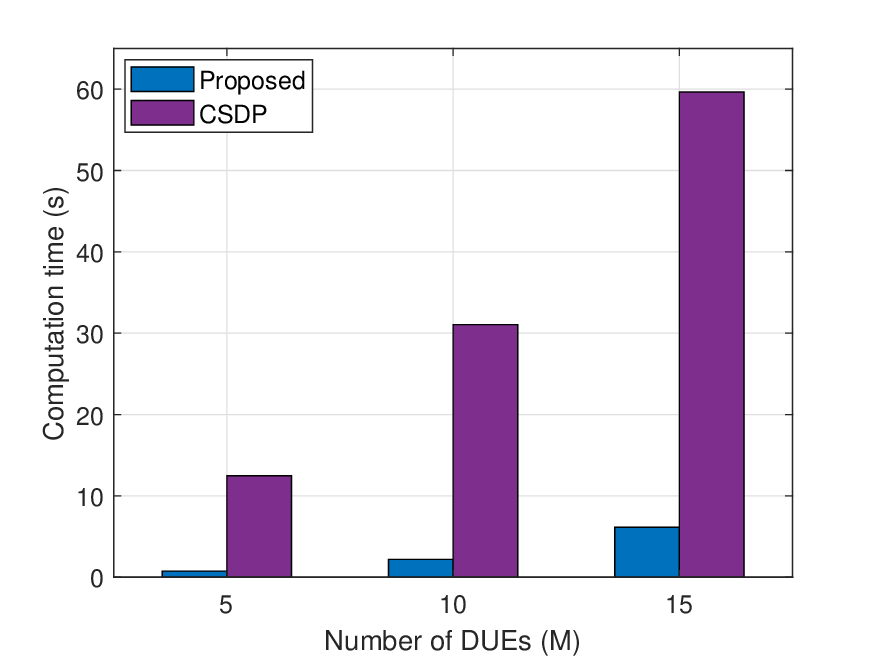}\\
			(a) Time vs. $N$ ($P_t=1 W$) & (b) Time vs. $M$ ($N=16$)\\
		\end{tabular}
	\end{minipage}
	\caption{Variation of computation time.}\label{ti}
\end{figure}

Finally, we compare the time complexity of the proposed consensus ADMM-based distributed cooperative algorithm with that of the SDP algorithm based on the interior point method in Fig. \ref{ti}. The results show that the proposed algorithm has better time efficiency with different TRIS elements and DUEs, and a 92.7\% reduction in complexity compared to the SDP algorithm based on the interior point method, and this feature is more suitable for future complex wireless networks.

\section{Conclusions}
In this paper, we have investigated cooperative ISAC network design through CUEs with RSMA. In particular, CUEs employ the novel transmissive RIS transceiver to realize cooperative sensing and communication in the form of low-cost and low-power consumption. Based on the network characteristics, a distributed cooperative ISAC algorithm has been designed, which utilizes consensus ADMM as a framework to achieve beamforming and user scheduling with low complexity. The results show that the proposed architecture outperforms traditional transceivers and enhances network coverage and the understanding of wireless environment. In addition, the complexity of the algorithm is much less than that of the baseline algorithm as the number of users and the number of TRIS elements increase, which is more applicable to the design of future wireless networks.

\bibliographystyle{IEEEtran}
\bibliography{IEEEabrv,reference}

%
%
%
%
%
%
%
%

\end{document}